%% file: birdtrack-lessons.tex
\documentclass[a4paper,12pt]{article}
\usepackage{amsmath,amsthm,amsfonts,amssymb,amscd}
\usepackage[colorlinks=true,linkcolor=blue,citecolor=blue,urlcolor=blue]{hyperref}
\usepackage[dvipsnames]{xcolor} 
\usepackage{framed,graphicx}
\usepackage[numbers,square]{natbib}
\usepackage{times}
\usepackage{framed}
\usepackage{lmodern}
\usepackage[french,english]{babel}
\usepackage[utf8]{inputenc}
\usepackage[T1]{fontenc}
\usepackage{dsfont}
\usepackage{cancel}
\usepackage[left=2.2cm, right=2.2cm, top=2.3cm , bottom = 2.2cm, headheight=14pt, headsep=30pt]{geometry}
\usepackage{setspace}
\usepackage{color}
\usepackage{fancyhdr, lastpage}
\usepackage{pifont} 
\usepackage{enumitem}
\usepackage{titlesec}
\include{macro}

\titlespacing\section{0pt}{12pt plus 4pt minus 2pt}{2pt plus 2pt minus 2pt}
\titlespacing\subsection{0pt}{12pt plus 4pt minus 2pt}{2pt plus 2pt minus 2pt}
\titlespacing\subsubsection{0pt}{2pt plus 0pt minus 2pt}{2pt plus 0pt minus 2pt}

\usepackage{tocloft}
\colorlet{shadecolor}{blue!8}
\theoremstyle{definition}

\newtheorem{exer}{Exercise}

\def \drawqua (#1) {
\draw[postaction={decorate},decoration={markings,mark=at position #1 with {\large\arrow{>}}}] 
}
\def \drawanti (#1) {
\draw[postaction={decorate},decoration={markings,mark=at position #1 with {\large\arrow{<}}}] 
}

\usepackage{tikz}
\usetikzlibrary{patterns}
\usetikzlibrary{decorations.pathmorphing,decorations.markings,trees,calc,shapes.geometric}
\tikzset{
  qua/.style={postaction={decorate},decoration={markings,mark=at position .55 with {\large\arrow{>}}}},
  anti/.style={postaction={decorate},decoration={markings,mark=at position .55 with {\large\arrow{<}}}},
  glu/.style={decorate,decoration={coil,amplitude=1.4pt, segment length=2.6pt}} ,
  glu2/.style={draw = blue} ,
  vertical align/.style={baseline=-.5*(height("$+$")-depth("$+$"))},
  every picture/.style={vertical align}
}
\newcommand{\tikeq}[1]{\vcenter{\hbox{\begin{tikzpicture}#1\end{tikzpicture}}}}
\newcommand{\tikeqbis}[1]{\hbox{\begin{tikzpicture}#1\end{tikzpicture}}}

\newenvironment{psmallmatrix}
{\left(\begin{smallmatrix}}
	{\end{smallmatrix}\right)}

\begin{document}

\title{Introduction to color in QCD: \\ Initiation to the birdtrack pictorial technique\footnote{Lectures given at the \href{https://indico.cern.ch/event/1183874/}{6th Chilean School of High Energy Physics}\! , Valparaiso, January 16-19, 2023.}}

\author{St\'ephane Peign\'e \\  
\small SUBATECH UMR 6457 (IMT Atlantique, Universit\'e de Nantes, IN2P3/CNRS) \\ 
\small 4 rue Alfred Kastler, 44307 Nantes, France \\
\small stephane.peigne@subatech.in2p3.fr}
\date{}

\maketitle
\begin{abstract}
These lectures are an elementary introduction to the "birdtrack" color pictorial technique, a useful tool in QCD. We review the basic rules, discuss color conservation and infinitesimal color rotations, learn how to project on partonic color states, how to derive their Casimir charges...~and at the same time learn a little bit of representation theory. 
\end{abstract}

\newpage 
\section*{Preamble}

The "birdtrack technique" aims to address the color structure of QCD in a pictorial way, drawing color graphs named birdtracks~\cite{Cvitanovic:2008zz} rather than writing mathematical symbols carrying color indices. We will consider $\sun$ (with $N \geq 3$) as the symmetry group of QCD, \ie, $N$ quark colors. For general $N$, the birdtrack technique is not more complicated than for "real QCD" ($N=3$). On the contrary, working with a general parameter $N$ usually brings a better understanding than working with fixed $N=3$. The birdtrack technique can be used as a handy tool in virtually any $\sun$ color calculation, whether it is evaluating simple color factors or addressing complex color structures that may otherwise seem out of reach.~I hope these lessons will give you a glimpse of the unspeakable and profound satisfaction that drawing birdtracks can bring. 

These courses do not require any prior knowledge (not even in QCD), except perhaps some notions of linear algebra, which will be recalled anyway. The lessons include a list of trivial and very simple exercises to keep the focus. Doing these exercises is mandatory to avoid superficial reading, and to learn how to use birdtracks in practice. 

To go beyond these introductory lessons, the birdtrack drawer is encouraged to consult the following helpful references. In Ref.~\cite{Dokshitzer:1995fv}, the color pictorial technique is illustrated by nice examples and exercises. Ref.~\cite{Keppeler:2017kwt} gives a pedagogical presentation of birdtracks and addresses a systematic method (different from that described in lessons~\ref{lesson3} and \ref{lesson4} of the present notes) to find the projectors on the color multiplets of multiparton systems. Ref.~\cite{Keppeler:2017kwt} also contains a substantial birdtrack bibliography. The color pictorial technique is a useful tool for QCD, but it has much broader applications.~The birdtrack lover should consult the valuable bible on the uses of birdtracks in group theory, Ref.~\cite{Cvitanovic:2008zz}.

\tableofcontents

\newpage

Color graphs are drawn with the Ti{\it k}Z {\LaTeX} package (Ti{\it k}Z version \pgfversion). 

\section{Pedestrian introduction}
\label{lesson1}

\subsection{The basic Lego bricks}
\label{sec:basic-legos}

The color structure of all interaction terms in the QCD Lagrangian can be expressed in terms of three elementary bricks, or "legos". Using these legos, the color structure of any QCD problem can in principle be worked out. The legos are defined and represented pictorially as 
\be \label{legos}
\tikeq{
  \draw (0,0) node[left] {$i$};
  \draw[qua] (0,0) -- (.5,0); 
  \draw[qua] (.5,0) -- (1,0);
  \draw (1,0) node[right] {$j$};
  \draw[glu] (.5,-0.5) -- (.5,0);
  \draw (.5,-0.5) node[below] {$a$};
} = (T^a)^{j}_{\ i} \quad ; \quad \quad
\tikeq{
  \draw (0,0) node[left] {$i$};
  \draw[anti] (0,0) -- (.5,0);
  \draw[anti] (.5,0) -- (1,0);
  \draw (1,0) node[right] {$j$};
  \draw[glu] (.5,-0.5) -- (.5,0);
  \draw (.5,-0.5) node[below] {$a$};
} = -(T^a)^{i}_{\ j} \quad ; \quad \quad  
\tikeq{
  \draw (0,0) node[left] {$c$};
  \draw[glu] (0,0) -- (.5,0)node{$\scriptstyle\bullet$}; 
  \draw[glu] (.5,0) -- (1,0);
  \draw (1,0) node[right] {$b$};
  \draw[glu](.5,-0.5) -- (.5,0);
  \draw (.5,-0.5) node[below] {$a$};
} =  -if_{abc}  \ . \hskip 1cm
\ee
As mentioned in the preamble, we consider the $\sun$ symmetry group ($N \geq 3$). Quark and antiquark color indices denoted by $i, j \ldots$ thus vary from 1 to $N$, and gluon indices denoted by $a, b\ldots$ vary from 1 to $N^2-1$. The matrices $T^a$ (with $a=1\ldots N^2-1$) are $N\times N$ Hermitian matrices ($(T^a)^\dagger \equiv {^t}(T^a)^* = T^a$), of zero trace,
\be
\tr{T^a} = 0 \, , 
\ee  
and normalized so that 
\be \label{trace-ab}
\tr{(T^a T^b)} = \frac{1}{2} \delta_{ab} \, .
\ee
The $f_{abc}$'s are the $\sun$ structure constants defining the Lie algebra of $\sun$, 
\be \label{Lie-fond}
\com{T^a}{T^b} = if_{abc} \,T^c \, , 
\ee
from which one can easily show that $f_{abc}$ is totally antisymmetric in $a, b, c$. 

To memorize the three legos \eq{legos}, one may view them as representing respectively quark, antiquark, and gluon scattering (the time arrow going from left to right) off an external gluon field (carrying the color index $a$) coupled {\it from below}, and remember that each lego corresponds to the $\sun$ {\it generator} of index $a$ in the representation of the scattered parton, namely, $T^a$ for a quark, $-T^a$ for an antiquark, and $t^a$ for a gluon, where the $(N^2-1) \times (N^2-1)$ matrices $t^a$ are defined by 
\be \label{ta}
(t^a)_{bc} = - if_{abc} \, .
\ee
Note however that the meaning of a generator (which will be recalled in lesson~\ref{lesson2}) is not really useful at this stage. 

An important feature of the rules \eq{legos} is that each lego is antisymmetric under the exchange of two lines. For instance, by exchanging any two lines of the first lego and rotating if needed the resulting graph in the mind's eye, one obtains a diagram looking like the second lego (antiquark scattering off a gluon coupled from below), which thus gets a minus sign. Let us also remark that in \eq{legos}, the color index of an outgoing quark (or incoming antiquark) is written by convention as an upper index, and that of an outgoing antiquark (or incoming quark) as a lower index. The usefulness of this convention will become clear below. 

In addition to \eq{legos}, we introduce the pictorial notation for Kronecker's of color indices:
\be \label{Kroneckers}
\tikeq{
   \draw (0,0) node[left] {$i$};
  \draw[qua](0,0) -- (1,0);
  \draw (1,0) node[right] {$j$};
} = \delta^{j}_{\ i} \quad ;  \quad \quad 
\tikeq{
   \draw (0,0) node[left] {$a$};
  \draw[glu](0,0) -- (1,0);
  \draw (1,0) node[right] {$b$};
} = \delta_{ab} \ . 
\ee
For those who have not yet had the chance to get acquainted with QCD or even Feynman diagrams, let us mention that in a "lego construction" (called color graph or birdtrack in what follows) built from the basic legos, the color index of an internal line is summed over all its possible values. Together with Einstein summation convention, according to which an index appearing twice in an expression is implicitly summed, we thus have, for an internal gluon line: 
\be
\tikeq{
		\draw[qua] (0,0)node[left]{$i$}  -- ++(0.5,0)coordinate(A);
		\draw[qua] (A) -- ++(0.5,0)node[right]{$j$} ;
		\draw[qua] (0,-.8)node[left]{$k$}  -- ++(0.5,0)coordinate(B);
		\draw[qua] (B) -- ++(0.5,0)node[right]{$l$} ;
		\draw[glu] (B) -- (A);
		\draw ($(A)+(0.25,-0.4)$) node{$\scriptstyle{(a)}$} ; 
		}
=   (T^a)^{j}_{\ i} (- T^a)^{l}_{\ k} = 	  
  \tikeq{
		\draw[qua] (0,0)node[left]{$i$}  -- ++(0.5,0)coordinate(A);
		\draw[qua] (A) -- ++(0.5,0)node[right]{$j$} ;
		\draw[glu] ($(A)+(0.,-0.4)$) -- (A);
		\draw ($(A)+(0.,-0.3)$) node[right]{$a$} ;
		\draw[qua] (0,-1)node[left]{$k$}  -- ++(0.5,0)coordinate(B);
		\draw[qua] (B) -- ++(0.5,0)node[right]{$l$} ;
		\draw[glu] (B) -- ($(B)+(0.,0.4)$) ; 
		\draw ($(B)+(0.,0.3)$) node[right]{$a$} ;
		} 
= \delta_{ab} 
\tikeq{
		\draw[qua] (0,0)node[left]{$i$}  -- ++(0.5,0)coordinate(A);
		\draw[qua] (A) -- ++(0.5,0)node[right]{$j$} ;
		\draw[glu] ($(A)+(0.,-0.4)$) -- (A);
		\draw ($(A)+(0.,-0.3)$) node[right]{$a$} ;
		\draw[qua] (0,-1)node[left]{$k$}  -- ++(0.5,0)coordinate(B);
		\draw[qua] (B) -- ++(0.5,0)node[right]{$l$} ;
		\draw[glu] (B) -- ($(B)+(0.,0.4)$) ; 
		\draw ($(B)+(0.,0.3)$) node[right]{$b$} ;
		} 
= 
\tikeq{
		\draw[qua] (0,0)node[left]{$i$}  -- ++(0.5,0)coordinate(A);
		\draw[qua] (A) -- ++(0.5,0)node[right]{$j$} ;
		\draw[glu] ($(A)+(0.,-0.4)$) -- (A);
		\draw ($(A)+(0.,-0.25)$) node[right]{$a$} ;
		\draw[qua] (0,-1.6)node[left]{$k$}  -- ++(0.5,0)coordinate(B);
		\draw[qua] (B) -- ++(0.5,0)node[right]{$l$} ;
		\draw[glu] (B) -- ($(B)+(0.,0.4)$) ; 
		\draw ($(B)+(0.,0.3)$) node[right]{$b$} ;
		\draw[glu] ($(B)+(0.,0.5)$) -- ($(A)+(0.,-0.5)$); 
		\draw ($(A)+(0.,-0.55)$) node[right]{$a$} ; 
		\draw  ($(B)+(0.,0.65)$) node[right]{$b$} ;
		} \ ,
\ee
and similarly, for an internal quark line: 
\be
\tikeq{
		\draw[glu] (0,0)node[left]{$a$}  -- ++(0.5,0)coordinate(A);
		\draw[anti] (A) -- ++(0.5,0)node[right]{$i$} ;
		\draw[glu] (0,-.8)node[left]{$b$}  -- ++(0.5,0)coordinate(B);
		\draw[qua] (B) -- ++(0.5,0)node[right]{$k$} ;
		\draw[anti] (B) -- (A);
		\draw ($(A)+(0.25,-0.4)$) node{$\scriptstyle{(j)}$} ; 
		}
=   (T^b)^{k}_{\ j} (T^a)^{j}_{\ i} = 	  
  \tikeq{
		\draw[glu] (0,0)node[left]{$a$}  -- ++(0.5,0)coordinate(A);
		\draw[anti] (A) -- ++(0.5,0)node[right]{$i$} ;
		\draw[anti] ($(A)+(0.,-0.4)$) -- (A);
		\draw ($(A)+(0.,-0.3)$) node[right]{$j$} ;
		\draw[glu] (0,-1)node[left]{$b$}  -- ++(0.5,0)coordinate(B);
		\draw[qua] (B) -- ++(0.5,0)node[right]{$k$} ;
		\draw[anti] (B) -- ($(B)+(0.,0.4)$) ; 
		\draw ($(B)+(0.,0.3)$) node[right]{$j$} ;
		} 
= \delta^{l}_{\ j}
  \tikeq{
		\draw[glu] (0,0)node[left]{$a$}  -- ++(0.5,0)coordinate(A);
		\draw[anti] (A) -- ++(0.5,0)node[right]{$i$} ;
		\draw[anti] ($(A)+(0.,-0.4)$) -- (A);
		\draw ($(A)+(0.,-0.3)$) node[right]{$j$} ;
		\draw[glu] (0,-1)node[left]{$b$}  -- ++(0.5,0)coordinate(B);
		\draw[qua] (B) -- ++(0.5,0)node[right]{$k$} ;
		\draw[anti] (B) -- ($(B)+(0.,0.4)$) ; 
		\draw ($(B)+(0.,0.3)$) node[right]{$l$} ;
		} 
= 
\tikeq{
		\draw[glu] (0,0)node[left]{$a$}  -- ++(0.5,0)coordinate(A);
		\draw[anti] (A) -- ++(0.5,0)node[right]{$i$} ;
		\draw[anti] ($(A)+(0.,-0.4)$) -- (A);
		\draw ($(A)+(0.,-0.25)$) node[right]{$j$} ;
		\draw[glu] (0,-1.8)node[left]{$b$}  -- ++(0.5,0)coordinate(B);
		\draw[qua] (B) -- ++(0.5,0)node[right]{$k$} ;
		\draw[anti] (B) -- ($(B)+(0.,0.4)$) ; 
		\draw ($(B)+(0.,0.3)$) node[right]{$l$} ;
		\draw[anti] ($(B)+(0.,0.6)$) -- ($(A)+(0.,-0.6)$); 
		\draw ($(A)+(0.,-0.65)$) node[right]{$j$} ; 
		\draw  ($(B)+(0.,0.75)$) node[right]{$l$} ;
		} \ .
\ee
Obviously, in pictorial notation summing over a repeated index amounts to connecting lines. The convention that a line with an arrow pointing out of (into) a vertex carries an upper (lower) index implies that for quarks and antiquarks, a repeated index always appears in both the upper and lower positions. In other words, the convention just serves to distinguish quarks and antiquarks and is consistent with the connection of quark lines respecting the direction of the arrow. 

In what follows, everything will be derived step by step, using only \eq{legos} and \eq{Kroneckers} for all knowledge. 

\subsection{First trivial rules}

The simplest way color appears in QCD is in the form of numbers called {\it color factors}. For example, in the calculation of the $qq \to qqg$ partonic cross section, one of the contributions is proportional to
\be \label{first-colour-factor}
\left( \ 
\tikeq{
		\draw[qua] (0,0)node[left]{$i$}  -- ++(0.4,0)coordinate(A);
		\draw (A) -- ++(0.4,0)coordinate(B);
		\draw (B) -- ++(0.4,0)coordinate(C);
		\draw[qua] (C) -- ++(0.4,0)node[right]{$j$} ;
		\draw[qua] (0,-.7)node[left]{$k$}  -- ++(0.4,0)coordinate(D);
		\draw (D) -- ++(0.4,0)coordinate(E);
		\draw (E) -- ++(0.4,0)coordinate(F);
		\draw[qua] (F) -- ++(0.4,0)node[right]{$l$} ;
		\draw[glu] (E) -- ($(B)+(0,.4)$)node{$\scriptstyle\bullet$};
		\draw[glu] (A) .. controls +(0,0.5) and +(-0.5,0) .. ($(A)+(1.2,.4)$)node[right]{$a$};   
} \ \right) 
\left( \ 
\tikeq{
                 \draw[qua] (0,0)node[left]{$i$} -- ++(0.4,0)coordinate(A);
		\draw (A) -- ++(0.4,0)coordinate(B);
		\draw (B) -- ++(0.4,0)coordinate(C);
		\draw[qua] (C) -- ++(0.4,0)node[right]{$j$};
		\draw[qua] (0,-.7)node[left]{$k$} -- ++(0.4,0)coordinate(D);
		\draw (D) -- ++(0.4,0)coordinate(E);
		\draw (E) -- ++(0.4,0)coordinate(F);
		\draw[qua] (F) -- ++(0.4,0)node[right]{$l$};
		\draw[glu] (E) -- (B);
		\draw[glu] (A) .. controls +(0,0.5) and +(-0.5,0) .. ($(A)+(1.2,.4)$)node[right]{$a$};  		
} \ \right)^* \ ,
\ee
where each factor is a {\it matrix element} in color space (\ie, a color graph with specified external indices), and every initial and final color index is implicitly summed. To calculate such color factors in a pictorial way, we first need the following rule: 
\begin{align}
& \emph{The complex conjugate of a color matrix element is obtained } \nn \\[-1mm]  
& \emph{pictorially by taking the mirror image of the associated graph} \hskip 5mm 
\label{rule1} \tag{R1} 
\\[-1mm] 
& \emph{and reversing the arrows of quark and antiquark lines.} \nn 
\end{align}

\bex
Check that this rule indeed holds for each of the three legos, and then show that it must be true for any matrix element built from these legos. 
\eex

Using rule \eq{rule1} and then summing over repeated indices by connecting lines, the color factor \eq{first-colour-factor} becomes
\be \label{first-colour-factor-2}
\tikeq{
\begin{scope}[scale=0.8]
		\draw (0,0) -- ++(1.6,0);
		\draw (0,-.7) -- ++(1.6,0);
		\draw (0,-1.3) -- ++(1.6,0);
		\draw (0,-2) -- ++(1.6,0);
		\drawanti(0.5) (0,0) arc (90:270:1.);
		\drawanti(0.5) (0,-.7) arc (90:270:.3);
                 \drawqua(0.5) (1.6,0) arc (90:-90:1.);
                 \drawqua(0.5) (1.6,-.7) arc (90:-90:.3);
		\draw[glu] (0.8,-.7) -- (0.8,.4)node{$\scriptstyle\bullet$};
	         \draw[glu] (0.8,-2) -- (.8,-1.3);
		\draw[glu] (0,0) .. controls +(0,0.5) and +(-0.5,0) .. ($(0,0)+(1.6,.4)$) arc (90:-90:1.4) .. controls +(-0.5,0) and +(0,-0.5) .. (0,-2); 
\end{scope} 
}  \ \ \ .
\vspace{2mm}
\ee
Such graphs can be evaluated with the help of simple pictorial rules. The simplest ones read 
\bea \label{eq:pict_trivial} 
\vspace{1mm}
\tikeq{
    \drawqua(0.55) (0,0) arc (180:0:.4);
    \draw (0,0) arc (-180:0:.4);
     } &=& 
 \tikeq{
    \drawqua(0.6) (0,0) arc (180:15:.4);
    \draw (0,0) arc (-180:-15:.4);
    \draw (.8,-.15) node[right] {$i$};
    \draw (.8,.25)  node[right] {$i$};
     } =  \delta^{i}_{\ i} = N \ , \\  \nn \\
\tikeq{    
    \draw[glu] (0,0) arc (0:360:.4);
    } 
&=&
\tikeq{    
    \draw[glu] (0,0) arc (15:345:.4);
    \draw (0,0.12) node[right] {$a$};
    \draw (0,-0.3) node[right] {$a$};
    } = \delta_{aa} = N^2-1  \ , \\ 
\tikeqbis{
    \draw[glu] (0,0) -- ++(.4,0);
    \drawanti(0.5) (.4,0) arc (180:-180:.4);
    \draw (0,0) node[above] {$a$};
    \draw (.47,.5) node {$\scriptstyle (i)$};
} &=& (T^a)^{i}_{\ i} = 0 \ , \\ 
\label{trace-norm-bird}
\tikeqbis{
    \draw[glu] (0,0) -- ++(.4,0);
    \draw[anti] (.4,0) arc (180:0:.4);
    \draw[qua] (.4,0) arc (180:360:.4);
    \draw[glu] (1.2,0) -- ++(.4,0);
    \draw (0,0) node[above] {$a$};
    \draw (.47,-.5) node {$\scriptstyle (j)$};
    \draw (1.6,0) node[above] {$b$};
    \draw (.47,.5) node {$\scriptstyle (i)$};
    } 
\hskip -3mm 
   &=& (T^a)^{j}_{\ i} (T^b)^{i}_{\ j} 
   = \frac{1}{2} 
\tikeqbis{
    \draw[glu] (0,0) -- ++(.8,0);
    \draw (0,0) node[above] {$a$};
    \draw (.8,0) node[above] {$b$};
}  \ .
\eea
In the next sections, more interesting rules are obtained using the pictorial representations of the Fierz identity and the Lie algebra. 

\subsection{Fierz identity}

In index notation, the Fierz identity reads 
\be
\delta^{i}_{\ j}\,\delta^{l}_{\ k} = \frac{1}{N} \, \delta^{i}_{\ k}\,\delta^{l}_{\ j} + 2 \, (T^a)^{i}_{\ k}\, (T^a)^{l}_{\ j} \, ,
\ee
which corresponds pictorially to 
\be \label{Fierz-index}
\tikeq{
    \draw[qua] (0,0)node[left] {$k$} -- ++(1,0)node[right] {$l$};
    \draw[anti] (0,.7)node[left] {$i$} -- ++(1,0)node[right] {$j$};
}\
=\ \frac{1}{N}\
\tikeq{
    \draw[qua] (0,0)node[left] {$k$} -- ++(0.2,0) -- ++(0,.7) -- ++(-0.2,0)node[left] {$i$};  
    \draw[anti] (0.7,0)node[right] {$l$} -- ++(-0.2,0) -- ++(0,.7) -- ++(0.2,0)node[right] {$j$};
}
+ 2\ 
\tikeq{
        \draw[anti] (0,.7)node[left]{$i$} -- ++(.3,-.35);
        \draw[qua] (0,0)node[left]{$k$} -- ++(.3,.35);
        \draw[glu] (.3,.35) -- ++(.7,0);
        \draw[anti] (1,.35) -- ++(.3,.35)node[right] {$j$} ;
        \draw[qua] (1,.35) -- ++(.3,-.35)node[right] {$l$} ;
} \ . 
\ee
For more comfort, we may write the latter equation by removing the external indices:
\vspace{1mm}
\be \label{Fierz-map}
\tikeq{ \draw[qua] (0,-0.35) -- ++(1,0);
  	  \draw[anti] (0,.35) -- ++(1,0); }
\ =\ \frac{1}{N}\
\tikeq{ \draw[qua] (0,-.35) -- ++(0.2,0) -- ++(0,.7) -- ++(-0.2,0);  
 	   \draw[anti] (0.7,-.35) -- ++(-0.2,0) -- ++(0,.7) -- ++(0.2,0); } 
+ 2\ 
\tikeq{ \draw[anti] (0,.35) -- ++(.3,-.35);
   	     \draw[qua] (0,-.35) -- ++(.3,.35);
   	     \draw[glu] (.3,0) -- ++(.7,0);
   	     \draw[anti] (1,0) -- ++(.3,.35);
   	     \draw[qua] (1,0) -- ++(.3,-.35) ; }  \ \ . 
\ee

It is important to note that in doing so, the mathematical meaning of the pictorial equation is changed: a color graph represents a {\it color matrix element} when the external indices are specified, or the corresponding {\it linear map} (between the vector spaces spanned by the objects carrying the initial and final indices) when external indices are removed. For instance, 
\be \label{Aklija}
\tikeq{
\begin{scope}[scale=0.7]
\drawqua(0.3) (-1.5,0.7)node[left]{$i$} -- (-0.6,0.7); 
\drawqua(0.85) (0.6,0.7) -- (1.5,0.7)node[right]{$k$}; 
\drawqua(0.3) (-1.5,0)node[left]{$j$} -- (-0.6,0); 
\drawqua(0.85) (0.6,0) -- (1.5,0)node[right]{$l$};  
\draw[glu] (-1.5,-0.7)node[left]{$a$} -- (-0.6,-0.7) ; 
\draw[fill=white] (0,0) circle (1); 
\draw (0,0) node{$A$} ; 
\end{scope}
} \equiv \ A^{kl}_{\  \ ija} 
\ee
are the matrix elements of the operator 
\be
\tikeq{
\begin{scope}[scale=0.7]
\drawqua(0.3) (-1.5,0.7) -- (-0.6,0.7); 
\drawqua(0.85) (0.6,0.7) -- (1.5,0.7); 
\drawqua(0.3) (-1.5,0) -- (-0.6,0); 
\drawqua(0.85) (0.6,0) -- (1.5,0);  
\draw[glu] (-1.5,-0.7) -- (-0.6,-0.7) ; 
\draw[fill=white] (0,0) circle (1); 
\draw (0,0) node{$A$} ; 
\end{scope}
}
\ \equiv \ A \, .
\ee

Note that the pictorial rule \eq{rule1} for complex conjugation of matrix elements corresponds to {\it Hermitian conjugation} at the operator level. Indeed, using \eq{Aklija} we have (using $A^* = {^t}A^\dagger$)
\be
( A^{kl}_{\  \ ija} )^* = ( A^\dagger)^{ija}_{\ \ \ kl} = 
\tikeq{
\begin{scope}[scale=0.7]
\drawqua(0.3) (-1.5,0.7)node[left]{$k$} -- (-0.6,0.7); 
\drawqua(0.85) (0.6,0.7) -- (1.5,0.7)node[right]{$i$}; 
\drawqua(0.3) (-1.5,0)node[left]{$l$} -- (-0.6,0); 
\drawqua(0.85) (0.6,0) -- (1.5,0)node[right]{$j$};  
\draw[glu] (0.6,-0.7) -- (1.5,-0.7)node[right]{$a$}; 
\draw[fill=white] (0,0) circle(1); 
\draw (0,0) node{$A^\dagger$} ; 
\end{scope}
} \, , 
\ee
but according to the rule \eq{rule1}, this must coincide with the graph obtained by taking the mirror image of \eq{Aklija} and reversing arrows. Thus, that transformation applied to the operator $A$ must give $A^\dagger$. 

The Fierz identity~\eq{Fierz-map} is thus a relation between linear maps of the vector space $V \otimes \overline{V}$ to itself, where $V$ and $\overline{V}$ are the quark and antiquark vector spaces, respectively. 
In particular, the l.h.s.~of~\eq{Fierz-map} is the identity operator: 
\vspace{1mm}
\be
\label{identity-qqbar}
\tikeq{ \Idqqbar }\ = \unit_{V \otimes \overline{V}} \, .
\ee

\bex 
\label{ex:fierz-proof}
Prove the Fierz identity~\eq{Fierz-map} by using the following linear algebra result: {\it If $E$ is a vector space of dimension $n$, $p_i$ ($i=1\ldots m$) are $m$ projectors ($p_i^2 = p_i$) such that $p_i \, p_j = 0$ for $i \neq j$ and $\sum_{i=1}^{m} {\rm rank}(p_i) = n$, then $\sum_{i=1}^{m} p_i = \unit_E$.} (Recall that ${\rm rank}(f) \equiv {\rm dim}[\img{f}]$.) Also use the fact that the rank of a projector is equal to its trace, and that the trace is simply obtained pictorially by connecting the initial and final lines carrying the same type of indices. For instance, the rank of the identity projector \eq{identity-qqbar} can be written as :
\be
{\rm rank}(\unit_{V \otimes \overline{V}}) = \tr{(\unit_{V \otimes \overline{V}})} = \tr{\left( \ \tikeq{\Idqqbar} \ \right)} = 
\tikeq{
 \draw[qua] (0,-0.2) -- (1,-0.2); 
 \draw (0,-0.2) .. controls +(-0.5,0) and +(-0.5,0) .. (0.5,-0.5) ; 
 \draw (0.5,-0.5) .. controls +(0.5,0) and +(0.5,0) .. (1,-0.2); 
 \draw[anti] (0,.1) -- (1,0.1);
 \draw (0,0.1) .. controls +(-0.5,0) and +(-0.5,0) .. (0.5,0.4) ; 
 \draw (0.5,0.4) .. controls +(0.5,0) and +(0.5,0) .. (1,0.1); 
} = N^2 \, , 
\ee
which coincides as it should with ${\rm dim}(V \otimes \overline{V})$. 
\eex

Let us now use the Fierz identity to derive simple pictorial rules. 

\bex
Multiply \eq{Fierz-index} by 
$\tikeq{\draw[qua] (0,0)node[left]{$i$} -- (0.6,0)node[right] {$j$};} = \delta^{j}_{\ i}$ and sum over repeated indices to show that 
\be \label{CF}
\tikeqbis{   \draw[qua] (0,0) -- (2,0); 
    		\draw[glu] (0.5,0) arc(180:0:.5);  } 
\ = \ C_F \ 
\tikeqbis{  \draw[qua] (0,0) -- (1,0); } \quad \quad \quad  (C_F = \frac{N^2-1}{2N}) \ .  
\ee
\eex

Anticipating the next lessons, let us note that Eq.~\eq{CF} can be written as $T^a T^a = C_F \unit_{V}$, where $T^a T^a$ is called the Casimir operator in the fundamental (quark) representation, and $C_F$ the (squared) quark color charge or simply "quark Casimir". (In any representation $R$, the Casimir operator $T^a(R) T^a(R)$ is proportional to the identity in that representation, as a consequence of Schur's lemma, see lesson~\ref{lesson3}.) 

\bex
Use the Fierz identity to obtain: 
\be \label{non-planar-bird}
\tikeqbis{\begin{scope}[yshift=0]
    \draw[qua] (0,0) -- (0.5,0); 
    \draw  (0.5,0) -- (1.5,0); 
    \draw[qua] (1.5,0) -- (2,0); 
    \draw[glu] (0.5,0) arc(180:0:.5); 
    \draw[glu] (1,-0.5) -- (1,0); 
\end{scope}} \ = \  - \frac{1}{2N} \ \ 
\tikeqbis{\begin{scope}[yshift=2]
    \draw[qua]  (0.4,0) -- (1,0); 
    \draw[qua] (1,0) -- (1.6,0); 
    \draw[glu] (1,-0.5) -- (1,0); 
\end{scope}} \ . 
\ee
\eex

\subsection{Lie algebra}

\subsubsection{Lie algebra in the fundamental representation}

The matrices $T^a$ (called the {\it generators} of $\sun$ in the fundamental representation, see lesson~\ref{lesson2}) satisfy the Lie algebra \eq{Lie-fond}, which can be expressed pictorially as
\be \label{Lie-fond-bird}
\tikeqbis{ \draw[glu] (0,0) -- (1,0); 
	   \drawqua(0.85) (1,0) .. controls +(0.2,0.2) and +(-0.5,0) .. (2,0.2); 
	 \drawanti(0.85) (1,0) .. controls +(0.2,-0.2) and +(-0.5,0) .. (2,-0.2); 
	 \draw[glu] (0.5,-0.5) -- (0.5,0)node{$\scriptstyle\bullet$};}
\ = \ 
\tikeqbis{ \draw[glu] (0,0) -- (1,0); 
	   \drawqua(0.85) (1,0) .. controls +(0.2,0.2) and +(-0.5,0) .. (2,0.2); 
	 \drawanti(0.85) (1,0) .. controls +(0.2,-0.2) and +(-0.5,0) .. (2,-0.2); 
	 \draw[glu] (1.5,-0.5) -- (1.5,0.19);}
\ + \ 
 \tikeqbis{ \draw[glu] (0,0) -- (1,0); 
	   \drawqua(0.85) (1,0) .. controls +(0.2,0.2) and +(-0.5,0) .. (2,0.2); 
	 \drawanti(0.85) (1,0) .. controls +(0.2,-0.2) and +(-0.5,0) .. (2,-0.2); 
	 \draw[glu] (1.5,-0.5) -- (1.5,-0.19);}
\ \ . 
\ee
\bex
Check it! 
\eex
The identity \eq{Lie-fond-bird} is the first example of "color conservation" that we encounter (see lesson~\ref{lesson2}). We will now use this identity to find new rules. 

\bex
\label{ex:CA}
Using {\it only} the pictorial rules derived so far, show that 
\be
\label{CA-bird}
\tikeqbis{   \draw[glu]  (0,0) -- (2,0); 
    		\draw[glu] (0.5,0)node{$\scriptstyle\bullet$} arc(180:0:.5) node{$\scriptstyle\bullet$}; } 
\ = \ N \ 
\tikeqbis{  \draw[glu] (0,0) -- (1,0); } \ . 
\ee
(Hint: to start, trade a gluon for a quark-antiquark pair by using \eq{trace-norm-bird}.)  
\eex

\bex
Starting from \eq{Lie-fond-bird}, show in two different ways the following rule: 
\be \label{planar-bird}
\tikeqbis{\begin{scope}[yshift=8]
    \draw[qua] (0,0) -- (0.5,0); 
    \draw  (0.5,0) -- (1.5,0); 
    \draw[qua] (1.5,0) -- (2,0); 
    \draw[glu] (1.5,0) arc(0:-180:.5); 
    \draw[glu] (1,-0.8) -- (1,-0.5)node{$\scriptstyle\bullet$}; 
\end{scope}} \ = \  \frac{N}{2} \  \ 
\tikeqbis{\begin{scope}[yshift=6]
    \draw[qua]  (0.4,0) -- (1,0); 
    \draw[qua] (1,0) -- (1.6,0); 
    \draw[glu] (1,-0.6) -- (1,0); 
\end{scope}} \ . 
\ee
(Hint: multiply \eq{Lie-fond-bird} either to the left by 
$\tikeqbis{\begin{scope}[yshift=2]
          \draw[glu] (0,0) -- (1,0); 
	 \draw[glu] (0.5,-0.3) -- (0.5,0)node{$\scriptstyle\bullet$};
	 \end{scope}}$\, , 
or to the right by 
$\tikeqbis{\begin{scope}[yshift=3]
   \draw[qua] (0,0) -- (0.5,0); 
   \draw[qua] (0.5,0) -- (1,0); 
   \draw[glu] (0.5,-0.3) -- (0.5,0);
\end{scope}}$\, , and sum over appropriate indices.) 
\eex

\subsubsection{Lie algebra in the adjoint representation}

A nice identity is the so-called Jacobi identity, 
\be \label{Jacobi-bird}
\tikeqbis{ \draw[glu] (0,0) -- (1,0)node{$\scriptstyle\bullet$}; 
	   \draw[glu] (1,0) .. controls +(0.2,0.2) and +(-0.5,0) .. (2,0.2); 
	 \draw[glu] (1,0) .. controls +(0.2,-0.2) and +(-0.5,0) .. (2,-0.2); 
	 \draw[glu] (0.5,-0.5) -- (0.5,0)node{$\scriptstyle\bullet$};}
\ = \ 
\tikeqbis{ \draw[glu] (0,0) -- (1,0)node{$\scriptstyle\bullet$}; 
	   \draw[glu] (1,0) .. controls +(0.2,0.2) and +(-0.5,0) .. (2,0.2); 
	 \draw[glu] (1,0) .. controls +(0.2,-0.2) and +(-0.5,0) .. (2,-0.2); 
	 \draw[glu] (1.5,-0.5) -- (1.5,0.19)node{$\scriptstyle\bullet$};}
\ + \ 
 \tikeqbis{ \draw[glu] (0,0) -- (1,0)node{$\scriptstyle\bullet$}; 
	   \draw[glu] (1,0) .. controls +(0.2,0.2) and +(-0.5,0) .. (2,0.2); 
	 \draw[glu] (1,0) .. controls +(0.2,-0.2) and +(-0.5,0) .. (2,-0.2); 
	 \draw[glu] (1.5,-0.5) -- (1.5,-0.19)node{$\scriptstyle\bullet$};}
\ \ ,
\ee
which is another manifestation of color conservation. 

\bex
Prove the Jacobi identity in two ways: 
\bi
\item[(i)] Verify that $\com{T^a}{\com{T^b}{T^c}}+\com{T^b}{\com{T^c}{T^a}}+\com{T^c}{\com{T^a}{T^b}}  = 0$ 
to infer the relation $f_{abe} f_{cde} + f_{bce} f_{ade} +f_{cae} f_{bde} = 0$, and check that the latter is equivalent to \eq{Jacobi-bird};  
\item[(ii)] The previous proof indicates that \eq{Jacobi-bird} is a direct consequence of the Lie algebra in the fundamental representation \eq{Lie-fond}, \ie\ of \eq{Lie-fond-bird}. Thus, it should be possible to prove the Jacobi identity \eq{Jacobi-bird} using only the pictorial rules derived so far. Try it! (Hint: use a similar start as in Exercise~\ref{ex:CA}.)
\ei
\eex

\bex
Check that the Jacobi identity is nothing but the expression of the $\sun$ Lie algebra in the adjoint representation, namely, 
\be
\label{Lie-adj}
\com{t^a}{t^b} = if_{abc} \, t^c \, , 
\ee
where the matrices $t^a$ are defined by \eq{ta}.   
\eex

Anticipating the next lessons, we note that since the $t^a$ matrices obey the Lie algebra and moreover satisfy $(t^a)^\dagger = t^a$ and $\tr{t^a} = 0$, they are $\sun$ generators (lesson~\ref{lesson2}) in a representation of dimension $N^2-1$ called the gluon or {\it adjoint} representation. The operator $t^a t^a$ is thus the {\it Casimir operator} (lesson~\ref{lesson3}) in the adjoint representation, and from {\it Schur's lemma} (lesson~\ref{lesson3}) we must have $t^a t^a = C_A \unit_{A}$ (with $A$ the gluon vector space of dimension $N^2-1$, and $C_A$ the "gluon Casimir"). In fact, we already found $t^a t^a$, since Eq.~\eq{CA-bird} reads (in index notation): 
\be
f_{acd} f_{bcd} = (t^c)_{ad} (t^c)_{db} = (t^c t^c)_{ab} = N \delta_{ab}  \ \ \Rightarrow \ \ t^c t^c = N \, \unit_{A} \, .
\ee
Therefore, the gluon Casimir is $C_A = N$.  

\bex
Use the Jacobi identity to obtain the rule: 
\be
\tikeqbis{    
    \draw[glu] (0,0) arc (180:-180:.4);
    \draw[glu] (-0.7,0) -- (0,0)node{$\scriptstyle\bullet$};
    \draw[glu] (0.7,0.3)node{$\scriptstyle\bullet$} -- ++ (0.65,0.35);
    \draw[glu] (0.7,-0.3)node{$\scriptstyle\bullet$} -- ++ (0.65,-0.35);
    } 
\ = \  \frac{N}{2} \  \ 
\tikeqbis{    
     \draw[glu] (-0.7,0) -- (0,0)node{$\scriptstyle\bullet$};
      \draw[glu] (0,0) -- ++ (0.65,0.35);
      \draw[glu] (0,0) -- ++ (0.65,-0.35);
}  \ \ . 
\ee
\eex

\subsection{Sum up}

In this lesson we have obtained the following set of simple pictorial rules  : 
\vspace{2.5mm}
\begin{align}
\tikeq{
    \drawqua(0.55) (0,0) arc (180:0:.4);
    \draw (0,0) arc (-180:0:.4);
     } =  N \ ; \quad 
\tikeq{    
    \draw[glu] (0,0) arc (0:360:.4);
    } 
= N^2-1 \ ; \quad 
\tikeqbis{
    \draw[glu] (0,0) -- ++(.4,0);
    \drawanti(0.5) (.4,0) arc (180:-180:.4);
} = 0 \ ; \quad 
\tikeqbis{
    \draw[glu] (0,0) -- ++(.4,0);
    \draw[anti] (.4,0) arc (180:0:.4);
    \draw[qua] (.4,0) arc (180:360:.4);
    \draw[glu] (1.2,0) -- ++(.4,0);
    } 
   = \frac{1}{2} \ 
\tikeqbis{
    \draw[glu] (0,0) -- ++(.8,0);} \hskip 0.8cm  & \nn \\ \nn \\ 
\tikeqbis{  \begin{scope}[scale=0.8,yshift=0] \draw[qua] (0,0) -- (2,0); 
    		\draw[glu] (0.5,0) arc(180:0:.5); \end{scope} } 
\ = \ C_F \ 
\tikeqbis{ \begin{scope}[scale=0.8,yshift=0] \draw[qua] (0,0) -- (1,0); \end{scope}} \ ; \quad 
\tikeqbis{ \begin{scope}[scale=0.8,yshift=0]  \draw[glu]  (0,0) -- (2,0); 
    		\draw[glu] (0.5,0)node{$\scriptstyle\bullet$} arc(180:0:.5) node{$\scriptstyle\bullet$}; \end{scope}} 
\ = \ C_A \ 
\tikeqbis{ \begin{scope}[scale=0.8,yshift=0]  \draw[glu] (0,0) -- (1,0); \end{scope}} \hskip 3cm & 
\label{rules:lesson1} \tag{R2} \\  \nn \\
\tikeqbis{\begin{scope}[scale=0.8,yshift=0]
    \draw[qua] (0,0) -- (0.5,0); 
    \draw  (0.5,0) -- (1.5,0); 
    \draw[qua] (1.5,0) -- (2,0); 
    \draw[glu] (0.5,0) arc(180:0:.5); 
    \draw[glu] (1,-0.5) -- (1,0); 
\end{scope}} \ = \  - \frac{1}{2N} \ \ 
\tikeqbis{\begin{scope}[scale=0.8,yshift=2]
    \draw[qua]  (0.4,0) -- (1,0); 
    \draw[qua] (1,0) -- (1.6,0); 
    \draw[glu] (1,-0.5) -- (1,0); 
\end{scope}} \ ; \quad 
\tikeqbis{\begin{scope}[scale=0.8,yshift=8]
    \draw[qua] (0,0) -- (0.5,0); 
    \draw  (0.5,0) -- (1.5,0); 
    \draw[qua] (1.5,0) -- (2,0); 
    \draw[glu] (1.5,0) arc(0:-180:.5); 
    \draw[glu] (1,-0.85) -- (1,-0.55)node{$\scriptstyle\bullet$}; 
\end{scope}} \ = \  \frac{N}{2} \  \ 
\tikeqbis{\begin{scope}[scale=0.8,yshift=6]
    \draw[qua]  (0.4,0) -- (1,0); 
    \draw[qua] (1,0) -- (1.6,0); 
    \draw[glu] (1,-0.6) -- (1,0); 
\end{scope}} \ ; \quad 
\tikeqbis{    \begin{scope}[scale=0.8,yshift=0]
    \draw[glu] (0,0) arc (180:-180:.4);
    \draw[glu] (-0.7,0) -- (0,0)node{$\scriptstyle\bullet$};
    \draw[glu] (0.7,0.3)node{$\scriptstyle\bullet$} -- ++ (0.65,0.35);
    \draw[glu] (0.7,-0.3)node{$\scriptstyle\bullet$} -- ++ (0.65,-0.35);
    \end{scope}} 
\ = \  \frac{N}{2} \  \ 
\tikeqbis{     \begin{scope}[scale=0.8,yshift=0]
     \draw[glu] (-0.7,0) -- (0,0)node{$\scriptstyle\bullet$};
      \draw[glu] (0,0) -- ++ (0.65,0.35);
      \draw[glu] (0,0) -- ++ (0.65,-0.35); 
\end{scope}} \hskip -5mm & \nn 
\end{align}

These rules will prove very useful in the next lessons. When one of the above loops appears as a sub-diagram in a color graph, the latter can be simplified by contracting the loop to its simpler r.h.s.~expression given in~\eq{rules:lesson1}. 

\bex
Using the rules \eq{rules:lesson1}, calculate the color factor \eq{first-colour-factor-2}. 
\eex
\vspace{-6mm}
\bex
Evaluate the color graph 
$\ \tikeqbis{ \begin{scope}[scale=1.5,yshift=0]
    \draw[glu] (0,0) arc (180:-180:.4);
    \draw[glu] (-0.4,0) -- (0,0)node{$\scriptstyle\bullet$};
    \draw[glu] (0.8,0)node{$\scriptstyle\bullet$} -- (1.3,0);
    \draw[glu] (0.1,-0.3)node{$\scriptstyle\bullet$} -- (0.7,0.3)node{$\scriptstyle\bullet$};
    \draw[glu] (0.1,0.3)node{$\scriptstyle\bullet$} -- (0.7,-0.3)node{$\scriptstyle\bullet$};
    \end{scope}}\ $. (Borrowed from Ref.~\cite{Dokshitzer:1995fv}.) 
\eex

\newpage

\section{Color conservation, color rotations and \texorpdfstring{$\bm{\sun}$}{} irreps}
\label{lesson2}

\subsection{Color conservation pictorially}
\label{sec:color-cons}

In lesson~\ref{lesson1}, we wrote the pictorial form of the Lie algebra, both in the fundamental and adjoint representation (see \eq{Lie-fond-bird} and \eq{Jacobi-bird}), 
\be \label{Lie-fond-bird-2}
\tikeqbis{ \draw[glu] (0,0) -- (1,0); 
	   \drawqua(0.85) (1,0) .. controls +(0.2,0.2) and +(-0.5,0) .. (2,0.2); 
	 \drawanti(0.85) (1,0) .. controls +(0.2,-0.2) and +(-0.5,0) .. (2,-0.2); 
	 \draw[glu] (0.5,-0.5) -- (0.5,0)node{$\scriptstyle\bullet$};}
\ = \ 
\tikeqbis{ \draw[glu] (0,0) -- (1,0); 
	   \drawqua(0.85) (1,0) .. controls +(0.2,0.2) and +(-0.5,0) .. (2,0.2); 
	 \drawanti(0.85) (1,0) .. controls +(0.2,-0.2) and +(-0.5,0) .. (2,-0.2); 
	 \draw[glu] (1.5,-0.5) -- (1.5,0.19);}
\ + \ 
 \tikeqbis{ \draw[glu] (0,0) -- (1,0); 
	   \drawqua(0.85) (1,0) .. controls +(0.2,0.2) and +(-0.5,0) .. (2,0.2); 
	 \drawanti(0.85) (1,0) .. controls +(0.2,-0.2) and +(-0.5,0) .. (2,-0.2); 
	 \draw[glu] (1.5,-0.5) -- (1.5,-0.19);}
\ \ ,
\ee
\be \label{Jacobi-bird-2}
\tikeqbis{ \draw[glu] (0,0) -- (1,0)node{$\scriptstyle\bullet$}; 
	   \draw[glu] (1,0) .. controls +(0.2,0.2) and +(-0.5,0) .. (2,0.2); 
	 \draw[glu] (1,0) .. controls +(0.2,-0.2) and +(-0.5,0) .. (2,-0.2); 
	 \draw[glu] (0.5,-0.5) -- (0.5,0)node{$\scriptstyle\bullet$};}
\ = \ 
\tikeqbis{ \draw[glu] (0,0) -- (1,0)node{$\scriptstyle\bullet$}; 
	   \draw[glu] (1,0) .. controls +(0.2,0.2) and +(-0.5,0) .. (2,0.2); 
	 \draw[glu] (1,0) .. controls +(0.2,-0.2) and +(-0.5,0) .. (2,-0.2); 
	 \draw[glu] (1.5,-0.5) -- (1.5,0.19)node{$\scriptstyle\bullet$};}
\ + \ 
 \tikeqbis{ \draw[glu] (0,0) -- (1,0)node{$\scriptstyle\bullet$}; 
	   \draw[glu] (1,0) .. controls +(0.2,0.2) and +(-0.5,0) .. (2,0.2); 
	 \draw[glu] (1,0) .. controls +(0.2,-0.2) and +(-0.5,0) .. (2,-0.2); 
	 \draw[glu] (1.5,-0.5) -- (1.5,-0.19)node{$\scriptstyle\bullet$};}
\ \ .
\ee
The latter equations can also be written by stretching the incoming gluon line to the final state (or the outgoing parton lines to the initial state). For instance, stretching the incoming gluon in \eq{Lie-fond-bird-2} to the final state, one obtains (due to the antisymmetry of the lego bricks under the exchange of two lines) 
\be \label{Lie-fond-bird-3}
0 \ = \ 
\tikeqbis{ \draw[glu] (1,0) .. controls +(0.2,0.2) and +(-0.5,0) .. (2,0.3); 
	   \drawqua(0.85) (1,0) -- (2,0); 
	 \drawanti(0.85) (1,0) .. controls +(0.2,-0.2) and +(-0.5,0) .. (2,-0.3); 
	 \draw[glu] (1.5,-0.6) -- (1.5,0.27)node{$\scriptstyle\bullet$};}
\ + \ 	 
\tikeqbis{ \draw[glu] (1,0) .. controls +(0.2,0.2) and +(-0.5,0) .. (2,0.3); 
	   \drawqua(0.85) (1,0) -- (2,0); 
	 \drawanti(0.85) (1,0) .. controls +(0.2,-0.2) and +(-0.5,0) .. (2,-0.3); 
	 \draw[glu] (1.5,-0.6) -- (1.5,0);}
\ + \ 
\tikeqbis{  \draw[glu] (1,0) .. controls +(0.2,0.2) and +(-0.5,0) .. (2,0.3); 
	   \drawqua(0.85) (1,0) -- (2,0); 
	 \drawanti(0.85) (1,0) .. controls +(0.2,-0.2) and +(-0.5,0) .. (2,-0.3); 
	 \draw[glu] (1.5,-0.6) -- (1.5,-0.27);}
\ .
\ee
Thus, for each lego the sum of gluon attachments to parton lines `before' and `after' the interaction vertex gives the same result, and this holds independently of the way the lego is represented in a time-ordered picture: either as a $1\to 2$ splitting (as in \eq{Lie-fond-bird-2} and \eq{Jacobi-bird-2}), or $2\to 1$, or $3\to 0$, or $0\to 3$ (as in \eq{Lie-fond-bird-3}). 

This trivially generalizes to any operator constructed from the lego bricks, leading to the pictorial representation of {\it color conservation}, 
\be
\label{color-cons-pict}
\tikeqbis{
\begin{scope}[scale=0.7]
\draw[qua] (-2.2,0.5) -- (-0.6,0.5); 
\draw[anti] (-2.2,0) -- (-0.6,0); 
\draw[glu] (-2.2,-0.5) -- (-0.6,-0.5) ; 
\draw[glu] (0.8,0.5) -- (2.3,0.5); 
\draw[glu] (0.8,0.17) -- (2.3,0.17);
\draw[glu] (0.8,-0.17) -- (2.3,-0.17);    
\draw[glu] (0.8,-0.5) -- (2.3,-0.5); 
\draw[fill=white,opacity=1] (0,0) circle (1); 
\draw (0,0) node{$A$} ;
\draw[glu] (-1.7,-1.3) -- (-1.7,-0.75); 
\draw[fill=white,opacity=0.5] (-1.7,0.7) .. controls +(-0.2,0) and +(-0.2,0) .. (-1.7,-0.75); 
\draw[fill=white,opacity=0.5] (-1.7,0.7) .. controls +(0.2,0) and +(0.2,0) .. (-1.7,-0.75);
\end{scope}
} 
\ \  = \ \ 
\tikeqbis{
\begin{scope}[scale=0.7]
\draw[qua] (-2,0.5) -- (-0.8,0.5); 
\draw[anti] (-2,0) -- (-0.8,0); 
\draw[glu] (-2,-0.5) -- (-0.8,-0.5) ; 
\draw[glu] (0.8,0.5) -- (2.3,0.5); 
\draw[glu] (0.8,0.17) -- (2.3,0.17);
\draw[glu] (0.8,-0.17) -- (2.3,-0.17);
\draw[glu] (0.8,-0.5) -- (2.3,-0.5); 
\draw[fill=white,opacity=1] (0,0) circle (1); 
\draw (0,0) node{$A$} ;
\draw[glu] (1.65,-1.3) -- (1.65,-0.75); 
\draw[fill=white,opacity=0.5] (1.65,0.7) .. controls +(-0.2,0) and +(-0.2,0) .. (1.65,-0.75); 
\draw[fill=white,opacity=0.5] (1.65,0.7) .. controls +(0.2,0) and +(0.2,0) .. (1.65,-0.75);
\end{scope}
} \ \ , 
\ee
where an ellipse crossed by a set of parton lines denotes the sum of all attachments of the "scattering gluon" to those lines. 
\bex
Although trivial, give a convincing proof of \eq{color-cons-pict}.  
\eex

\subsection{Color rotations}

\subsubsection{Finite \texorpdfstring{$\sun$}{} transformations}

The special unitary group $\sun$ is the Lie group of $N\times N$ unitary matrices with unit determinant, 
\be
U \in \sun \ \Leftrightarrow \ U U^\dagger  = \unit \ \ {\rm and} \ \ {\rm det} \, U = 1 \, . 
\ee
Any element of $\sun$ can be parametrized by 
\be
\label{SUN-param}
U(\alpha) = e^{i \alpha^a T^a} \, , 
\ee
where the matrices $T^a$ ($a=1\ldots N^2-1$) are the Hermitian matrices introduced in lesson~\ref{lesson1} (hence the name of $\sun$ {\it generators} for those matrices). The real parameters $\alpha^a$ may be viewed as the "angles" of the "color rotation" $U(\alpha)$.  
\bex
Check that the matrix \eq{SUN-param} indeed belongs to $\sun$. (In fact, the   
exponential parametrization \eq{SUN-param} generates {\it all} elements of $\sun$, see mathematics textbooks for a proof.) 
\eex

By construction, the QCD lagrangian is invariant under $\sun$ transformations or "color rotations". In order to address the color structure of QCD (in particular, to determine the invariant multiplets of a parton system), 
we first consider $\sun$ transformations of the quark, antiquark and gluon "color coordinates". 

\vspace{3mm}
\subsubsection*{Color rotations of quark and antiquark coordinates}

Let us start with quarks and antiquarks. 
Under a given color rotation $U(\alpha) \in \sun$, the quark coordinates (denoted by an {\it upper} index according to our initial convention, see section~\ref{sec:basic-legos}) transform as 
\be
\label{q-trans}
q^\prime = U \, q \ \Leftrightarrow \  q^{\prime \, i} = U^{i}_{\ j}  \, q^j \, .
\ee
When restricting to the color degree of freedom, antiquark coordinates are simply obtained from quark coordinates by complex conjugation. In the same color rotation of angles $\alpha^a$, antiquark coordinates thus transform as 
\be
\label{anti-color-rot-star}
q^{* \prime} = U^{*} \, q^{*} \ \Leftrightarrow \  (q^{* \prime})^{i} = (U^{*})^{i}_{\ j}  \, (q^{*})^j \, . 
\ee
A standard convention is to denote complex conjugation by moving quark and antiquark indices up and down, namely, 
\be
(q^{*})^{i}  \equiv q_{i} \ ; \ \ (U^{*})^{i}_{\ j} \equiv  U_{i}^{\ j} \, ,
\ee
a convention that we have implicitly used from the beginning by assigning {\it lower} color indices to antiquarks, see section~\ref{sec:basic-legos}. The transformation \eq{anti-color-rot-star} of antiquark coordinates is then written as
\be
q^{\prime}_{i} = U_{i}^{\ j}  \, q_j \, . 
\ee

To complement the above convention, any quantity transforming as quark (antiquark) coordinates is assigned an upper (lower) index. We readily verify that a product of the form $A^i  B_i$ (implicitly summed over  $i$) is $\sun$ invariant. Indeed, $(A^i  B_i)^\prime = U^{i}_{\ j} \, U_{i}^{\ k}  \, A^j  B_k = A^i  B_i$, since $U^{i}_{\ j} \, U_{i}^{\ k} = U^{i}_{\ j}  (U^*)^{i}_{\ k} = U^{i}_{\ j}  (U^\dagger)^{k}_{\ i} = (U^\dagger U)^{k}_{\ j} = \delta^{k}_{\ j}$. 

\vspace{2mm}
Under two successive color rotations of angles $\alpha^a$ and $\beta^b$, quark coordinates transform as
\be
\label{2rotations-quark}
q \ \mathop{\longrightarrow}^{\alpha} \  U(\alpha)\,  q \  \mathop{\longrightarrow}^{\beta} \  U(\beta) U(\alpha) \, q = U(\gamma(\alpha,\beta)) \, q \, . 
\ee
Indeed, since $\sun$ is a group, the product $U(\beta) U(\alpha)$ must coincide with an element of $\sun$ of angles $\gamma^c$, the latter being fully determined by $\alpha^a$ and $\beta^b$. 

\bex
\label{ex:BCH}
(To be done once in a lifetime.) 

Let us recall the Baker-Campbell-Hausdorff formula for the product of two exponentials of matrices, 
\be
\label{eq:BCH}
e^{X} \, e^{Y} = e^{X+Y+\frac{1}{2} \com{X}{Y} + \frac{1}{12} \left(\com{X}{\com{X}{Y}} - \com{Y}{\com{X}{Y}} \right) + \ldots } \, , 
\ee
where the dots stand for higher-order terms in $X$ and $Y$ (all being nested commutators of $X$ and $Y$). 
Using \eq{eq:BCH}, show that the angles $\gamma^a$ defined by \eq{2rotations-quark} are given by $\gamma^a(\alpha,\beta) = \alpha^a + \beta^a  +\frac{1}{2} f_{abc} \alpha^b \beta^c + \ldots$\,, and find the next term in the series.
\eex
This exercise illustrates that the structure of $\sun$ (with respect to the multiplication law) is fully determined by the $\sun$ Lie algebra \eq{Lie-fond}. 

\vspace{3mm}
\subsubsection*{Color rotations of gluon coordinates}

How should the $N^2-1$ gluon coordinates $\Phi^a$ transform under a color rotation of angles $\alpha^a$, when the $N$ quark coordinates transform with the matrix $U(\alpha)$? For each $U(\alpha)$ acting in quark space, we must find a corresponding $(N^2-1) \times (N^2-1)$ matrix ${\tilde U}(\alpha)$ acting in gluon space, in such a way that the "representation" $U(\alpha) \to {\tilde U}(\alpha)$ preserves the group structure. Indeed, in the successive rotations of angles $\alpha$ and $\beta$, the gluon coordinates become
\be
\Phi \ \mathop{\longrightarrow}^{\alpha} \  {\tilde U}(\alpha)\,  \Phi \  \mathop{\longrightarrow}^{\beta} \  {\tilde U}(\beta) {\tilde U}(\alpha) \, \Phi \, , 
\ee
but for consistency with \eq{2rotations-quark}, the same result should be obtained by a single rotation of angles $\gamma(\alpha,\beta)$, represented by ${\tilde U}(\gamma(\alpha,\beta))$ when acting in gluon space. We thus need
\be
\label{group-homomorphism}
{\tilde U}(\beta) {\tilde U}(\alpha) = {\tilde U}(\gamma(\alpha,\beta)) \, ,
\ee
with the same function $\gamma(\alpha,\beta)$ as derived in Exercise~\ref{ex:BCH}. 

It is clear that \eq{group-homomorphism} will be satisfied by the matrices 
\be
\label{SUN-param-adj}
{\tilde U}(\alpha) = e^{i \alpha^a {\tilde T}^a} \, ,
\ee
provided one can find $(N^2-1) \times (N^2-1)$ matrices ${\tilde T}^a$ ($a=1\ldots N^2-1$) having the same Lie algebra as the $T^a$'s, namely, 
\be 
\com{{\tilde T}^a}{{\tilde T}^b} = if_{abc} \,{\tilde T}^c \, .
\ee
We know from lesson~\ref{lesson1} that such matrices exist: the matrices $t^a$ defined by \eq{ta} satisfy \eq{Lie-adj}. 

\vspace{2mm}
A few remarks: 
\vspace{-2mm}
\bi
\item[\ding{118}] The set of matrices $U(\alpha) = e^{i \alpha^a T^a}$ (\ie, the $\sun$ group itself) acting on the quark and ${\tilde U}(\alpha) = e^{i \alpha^a t^a}$ acting on the gluon are respectively called the {\it fundamental} and {\it adjoint} $\sun$ representations.
\item[\ding{118}] The adjoint representation is real: ${\tilde U}(\alpha)^* = e^{-i \alpha^a (t^a)^*} = e^{i \alpha^a t^a} = {\tilde U}(\alpha)$. 
\item[\ding{118}] If there are $d_R \times d_R$ matrices $T^a(R)$ ($a=1\ldots N^2-1$) satisfying the $\sun$ Lie algebra, $\com{T^a(R)}{T^b(R)} = if_{abc} \,T^c(R)$, the matrices $U_R(\alpha) = e^{i \alpha^a T^a(R) }$ define an $\sun$ representation of dimension $d_R$, acting on objects with $d_R$ components while preserving the group structure. The $T^a(R)$'s are the $\sun$ generators in the representation $R$. 
\item[\ding{118}] For $N>2$, $\sun$ representations do not exist for any dimension $d_R$. For $N=3$, the possible dimensions are $d_R = 1, 3, 6, 8, 10, 15\ldots$ 
\item[\ding{118}] When there is no risk of confusion, an $\sun$ (irreducible) representation is labelled by its dimension in the case $N=3$. For instance, the fundamental and adjoint $\sun$ representations are denoted by $R={\bf 3}$ and $R={\bf 8}$, with generators $T^a({\bf 3}) = T^a$ and $T^a({\bf 8}) = t^a$. 
\item[\ding{118}] The antiquark transforms under the complex conjugate of the fundamental representation, denoted by $R={\bf \bar{3}}$ and given by the set of $N \times N$ matrices $U(\alpha)^* \equiv e^{i \alpha^a T^a({\bf \bar{3}})}$, with generators $T^a({\bf \bar{3}}) = - (T^a)^*$. Although the representations ${\bf 3}$ and ${\bf \bar{3}}$ have the same dimension $N$, they are not equivalent (for $N>2$), \ie, $U(\alpha)$ and $U(\alpha)^*$ are not related by a change of basis, and thus describe the transformations of different objects. 
\ei

\subsubsection{Infinitesimal color rotations}

$\sun$ is a Lie group for which infinitesimal transformations capture most of the group structure~\cite{Georgi:2000vve}. In particular, it is sufficient to consider infinitesimal transformations to highlight $\sun$ representations (see section~\ref{sec:sun-irreps}). 

Let us consider a color rotation of infinitesimal angles $\delta \alpha^a$.
According to \eq{SUN-param} and \eq{q-trans}, the quark transforms as 
\be
q^{\prime \, i} = q^i + i \delta \alpha^a \,  (T^a)^{i}_{\ j}  \, q^j \, ,
\ee
from which the transformation of the antiquark directly follows (take the complex conjugate, and recall that $(T^a)^* = {^t}T^a$): 
\be
q^{\prime}_{i} = q_i - i \delta \alpha^a  \, q_j \, (T^a)^{j}_{\ i}  \,  . 
\ee
Using \eq{SUN-param-adj}, the gluon transforms as:
\be
\Phi^{\prime \, b} = \Phi^b + i \delta \alpha^a \,  (t^a)_{bc}  \, \Phi^c \, .
\ee

The {\it infinitesimal shifts} of the quark, antiquark and gluon coordinates thus read
\bea
\delta q^i &\equiv& q^{\prime \, i} - q^i \ = \  i  \delta \alpha^a  \ \ 
\tikeq{
  \draw[qua] (0,0) -- (.5,0); 
  \draw[qua] (.5,0) -- (1,0);
  \draw (1,0) node[right] {$i$};
  \draw[glu] (.5,-0.5) -- (.5,0);
  \draw (.5,-0.5) node[right] {$a$};
  \draw[fill=white,opacity=1] (-0.1,0) circle (0.1); 
}  \ , \\
\delta  q_{i} &\equiv& q^{\prime}_{i} - q_i  \ = \   i \delta \alpha^a \ \ 
\tikeq{
  \draw[anti] (0,0) -- (.5,0);
  \draw[anti] (.5,0) -- (1,0);
  \draw (1,0) node[right] {$i$};
  \draw[glu] (.5,-0.5) -- (.5,0);
  \draw (.5,-0.5) node[right] {$a$};
  \draw[fill=white,opacity=1] (-0.1,0) circle (0.1); 
} \ ,  \\ 
\delta  \Phi^b &\equiv& \Phi^{\prime \, b} - \Phi^b \ = \  i  \delta \alpha^a  \ \ 
\tikeq{
  \draw[glu] (0,0) -- (.5,0); 
  \draw[glu] (.5,0) -- (1,0);
  \draw (1,0) node[right] {$b$};
  \draw[glu] (.5,-0.5) -- (.5,0);
  \draw (.5,-0.5) node[right] {$a$};
  \draw[fill=white,opacity=1] (-0.1,0) circle (0.1); 
}  \ , 
\eea
where we introduced the pictorial notation for coordinates: 
\be
\tikeq{
\begin{scope}[scale=1.2]
\draw[qua] (0,0) -- (.5,0)node[right] {$j$}; 
\draw[fill=white,opacity=1] (-0.1,0) circle (0.1); 
\end{scope}
}
\equiv q^j  \ \  ; \ \ \ \  
\tikeq{
\begin{scope}[scale=1.2]
\draw[anti] (0,0) -- (.5,0)node[right] {$j$}; 
\draw[fill=white,opacity=1] (-0.1,0) circle (0.1); 
\end{scope}
}
\equiv q_j   \ \  ; \ \ \ \  
\tikeq{
\begin{scope}[scale=1.2]
\draw[glu] (0,0) -- (.5,0)node[right] {$c$}; 
\draw[fill=white,opacity=1] (-0.1,0) circle (0.1); 
\end{scope}
}
\equiv \Phi^c  \  . 
\ee

Our basic legos \eq{legos} are defined as the $\sun$ generators in the quark, antiquark and gluon representations. Up to the factor $i \delta \alpha^a$, the legos are thus nothing but the infinitesimal shift of the corresponding parton coordinates. 
In other words, in a color rotation of angles $\delta \alpha^a$ the infinitesimal shift of parton coordinates is obtained pictorially (up to the factor $i \delta \alpha^a$) by attaching a gluon of color $a$ from below to the corresponding line. 

\vspace{3mm}
Let us now rewrite the "color conservation identity" \eq{color-cons-pict} as 
\be
\label{color-cons-2}
 i  \delta \alpha^a  \ \ 
\tikeqbis{
\begin{scope}[scale=1]
\draw[qua] (-2.2,-0.3) -- (-0.6,-0.3); \draw[fill=white,opacity=1] (-2.3,-0.3) circle (0.1); 
\draw[anti] (-2.2,-0.6) -- (-0.6,-0.6); \draw[fill=white,opacity=1] (-2.3,-0.6) circle (0.1); 
\draw[glu] (-2.2,-0.9) -- (-0.6,-0.9) ;\draw[fill=white,opacity=1] (-2.3,-0.9) circle (0.1); 
\draw[glu] (-2.2,0.) -- (-0.6,0.) ;\draw[fill=white,opacity=1] (-2.3,0) circle (0.1); 
\draw[glu] (-2.2,0.3) -- (-0.6,0.3) ;\draw[fill=white,opacity=1] (-2.3,0.3) circle (0.1); 
\draw[glu] (-2.2,0.6) -- (-0.6,0.6) ;\draw[fill=white,opacity=1] (-2.3,0.6) circle (0.1); 
\draw[glu] (-2.2,0.9) -- (-0.6,0.9) ;\draw[fill=white,opacity=1] (-2.3,0.9) circle (0.1); 
\draw[fill=white,opacity=1] (0,0) circle (1.2); 
\draw (0,0) node{$A$} ;
\draw[glu] (-1.7,-1.5) -- (-1.7,-1.1);  
\draw (-1.7,-1.5) node[right] {$a$};
\draw[fill=white,opacity=0.5] (-1.7,1.1) .. controls +(-0.2,0) and +(-0.2,0) .. (-1.7,-1.1); 
\draw[fill=white,opacity=0.5] (-1.7,1.1) .. controls +(0.2,0) and +(0.2,0) .. (-1.7,-1.1);
\end{scope}
} 
\ = \ 0 \ \ .
\ee
In \eq{color-cons-2}, the sum of the infinitesimal shifts is obviously the infinitesimal shift of the incoming multi-parton state. Thus, a parton system which is fully contracted over parton color indices is $\sun$ invariant. Such a system is called a color singlet state. Color conservation (expressed pictorially as \eq{color-cons-pict}) is equivalent to the $\sun$ invariance of color singlet systems. 

Note that if we do not contract with external parton coordinates, the identity \eq{color-cons-2} reads  
\be
\label{color-cons-tensor}
\tikeqbis{
\begin{scope}[scale=1]
\draw[qua] (-2.2,-0.3)node[left]{{\scriptsize $i$}}  -- (-0.6,-0.3); 
\draw[anti] (-2.2,-0.6)node[left]{{\scriptsize $j$}}  -- (-0.6,-0.6); 
\draw[glu] (-2.2,-0.9) -- (-0.6,-0.9) ;
\draw[glu] (-2.2,0.) -- (-0.6,0.) ;
\draw[glu] (-2.2,0.3) -- (-0.6,0.3) ;
\draw[glu] (-2.2,0.6)node[left]{{\scriptsize $c$}}  -- (-0.6,0.6) ;
\draw[glu] (-2.2,0.9)node[left]{{\scriptsize $b$}}  -- (-0.6,0.9) ;
\draw[line width=0.4mm,dotted,color=black!80,opacity =1]  (-2.43,0.37) -- (-2.43,-0.1); 
\draw[line width=0.4mm,dotted,color=black!80,opacity =1]  (-2.43,-0.8) -- (-2.43,-1); 
\draw[fill=white,opacity=1] (0,0) circle (1.2); 
\draw (0,0) node{$A$} ;
\draw[glu] (-1.7,-1.5) -- (-1.7,-1.1); 
\draw[fill=white,opacity=0.5] (-1.7,1.1) .. controls +(-0.2,0) and +(-0.2,0) .. (-1.7,-1.1); 
\draw[fill=white,opacity=0.5] (-1.7,1.1) .. controls +(0.2,0) and +(0.2,0) .. (-1.7,-1.1);
\end{scope}
} 
\ = \ 0 \, ,
\ee
with specified external indices $b, c,\ldots, i, j, \ldots$ Let us view the object carrying those indices, $A^{bc \ldots \ \ \ j \ldots}_{\ \ \ \ i \ldots}$, as an $\sun$ tensor, thus transforming under $\sun$ as the product of parton coordinates $\Phi^b \Phi^c \ldots q_i \ldots q^j \ldots$ Eq.~\eq{color-cons-tensor} gives an alternative formulation of color conservation, namely: all $\sun$ tensors (constructed from the basic legos) are in fact  $\sun$ {\it invariant} tensors. 

\bex Check explicitly that the tensors 
\be
\tikeq{
   \drawqua(0.3) (0,0.2) node[left]{{\scriptsize $i$}}  .. controls +(0.5,0) and +(0,0.2) .. (1,0) ;
  \drawanti(0.3) (0,-0.2) node[left]{{\scriptsize $j$}}  .. controls +(0.5,0) and +(0,-0.2) .. (1,0) ;
} = \delta^{j}_{\ i} \ \ , \ \  
\tikeq{
   \draw[glu] (0,0.2) node[left]{{\scriptsize $b$}}  .. controls +(0.5,0) and +(0,0.2) .. (1,0)  .. controls +(0,-0.2) and +(0.5,0) .. (0,-0.2)node[left]{{\scriptsize $c$}} ;
} = \delta^{bc} \ \ ,
\ee
are invariant under finite color rotations. 
\eex

\bex
Express the $\sun$ invariance under finite color rotations of the tensor
\be
\tikeq{
   \drawqua(0.3) (0,0.3) node[left]{{\scriptsize $i$}}  .. controls +(0.5,0) and +(0,0.2) .. (1,0) ;
  \drawanti(0.3) (0,-0.3) node[left]{{\scriptsize $j$}}  .. controls +(0.5,0) and +(0,-0.2) .. (1,0) ;
   \draw[glu] (0,0) node[left]{{\scriptsize $a$}} --  (1,0) ;} = (T^a)^{j}_{\ i}
\ee
to obtain the relation
\be
\tilde{U}_{bc} = 2 \tr{(T^b U T^c U^\dagger)} \, , 
\ee
which determines the matrix elements $\tilde{U}_{bc}$ of a color rotation in the adjoint representation in terms of its fundamental representation $U$. 
\eex

\subsection{\texorpdfstring{$\sun$}{} irreducible representations}
\label{sec:sun-irreps}

Using the pictorial expression of color conservation and infinitesimal color rotations allows one to address $\sun$ irreducible representations in a rather intuitive way. 

Consider a multi-parton system spanning a color vector space $E$ of dimension $n$, and suppose we have at disposal $m$ projectors $\proj_i$ constructed from the basic legos and satisfying the conditions $\proj_i \cdot \proj_j = 0$ for $i \neq j$ and $\sum_{i=1}^{m} {\rm rank}(\proj_i) = n$, implying the completeness relation $\sum_{i=1}^{m} \proj_i = \unit_E$. 
(An explicit case was given in lesson~\ref{lesson1} when proving the Fierz identity, see Exercise \ref{ex:fierz-proof}.) We also suppose the projectors to be Hermitian, $\proj_i^\dagger = \proj_i$. 

Let us apply an infinitesimal color rotation to the parton state (for the argument it is sufficient to keep only the infinitesimal shift and drop the factor $i \delta \alpha^a$), and then insert on the left and right the completeness relation : 
\be 
\tikeqbis{
\begin{scope}[scale=1]
\draw[qua] (0,0.5) -- (1,0.5); 
\draw[glu] (0,0) -- (1,0) ; 
\draw[anti] (0,-0.5) -- (1,-0.5); 
\end{scope}
} \ \ \ 
\mathop{\longrightarrow}^{\rm{inf.}}_{\rm{shift}} \ \ \ 
\tikeqbis{
\begin{scope}[scale=1]
\draw (0,0.5) -- (1,0.5); \draw[qua] (0,0.5) -- (0.4,0.5); \draw[qua] (0.6,0.5) -- (1,0.5); 
\draw[glu] (0,0) -- (1,0) ; 
\draw (0,-0.5) -- (1,-0.5); \draw[anti] (0,-0.5) -- (0.4,-0.5); \draw[anti] (0.6,-0.5) -- (1,-0.5); 
\draw[glu] (0.5,-1) -- (0.5,-0.6); 
\draw[fill=white,opacity=0.5] (0.5,0.6) .. controls +(-0.15,0) and +(-0.15,0) .. (0.5,-0.6); 
\draw[fill=white,opacity=0.5] (0.5,0.6) .. controls +(0.15,0) and +(0.15,0) .. (0.5,-0.6);
\end{scope}
}  \ = \ \sum_{i,j} \ 
\tikeqbis{
\begin{scope}[scale=1]
\draw (-1.7,0.5) -- (2.7,0.5);
 \draw[qua] (-1.7,0.5) -- (-1.2,0.5);  \draw[qua] (2.2,0.5) -- (2.7,0.5); 
 \draw[qua] (0,0.5) -- (0.2,0.5); \draw[qua] (0.8,0.5) -- (1,0.5); 
\draw[glu] (-1.7,0) -- (2.7,0) ; 
\draw (-1.7,-0.5) -- (2.7,-0.5); 
\draw[anti] (-1.7,-0.5) -- (-1.2,-0.5);  \draw[anti] (2.2,-0.5) -- (2.7,-0.5);
\draw[anti] (0,-0.5) -- (0.2,-0.5); \draw[anti] (0.8,-0.5) -- (1,-0.5); 
\draw[glu] (0.5,-1) -- (0.5,-0.6); 
\draw[fill=white,opacity=0.5] (0.5,0.6) .. controls +(-0.15,0) and +(-0.15,0) .. (0.5,-0.6); 
\draw[fill=white,opacity=0.5] (0.5,0.6) .. controls +(0.15,0) and +(0.15,0) .. (0.5,-0.6);
\draw[fill=white,opacity=1] (-0.7,0) circle (0.6); \draw (-0.7,0) node{$\proj_i$} ;
\draw[fill=white,opacity=1] (1.7,0) circle (0.6); \draw (1.7,0) node{$\proj_j$} ;
\draw[line width=0.4mm,dashed,color=black!40,opacity =1]  (0.24,-0.9) -- (0.24,0.65);
\draw (0.27,0.55) node[above]{{\scriptsize $\img{\proj_i}$}} ;
\draw[line width=0.4mm,dashed,color=black!40,opacity =1] (2.55,-0.9) -- (2.55,0.65); 
\draw (2.6,0.54) node[above]{{\scriptsize $\img{\proj_j}$}} ;
\end{scope}
} 
\ , 
\ee
where a dashed vertical line indicates to which subspace the corresponding intermediate multi-parton state belongs (here $\img{\proj_i}$ or $\img{\proj_j}$). 

Using color conservation and $\proj_i \cdot \proj_j = 0$ for $i \neq j$, only the terms with $i=j$ remain in the double sum. As a consequence, the image space $\img{\proj_i}$ of the projector $\proj_i$ is invariant under any infinitesimal color rotation, and thus under $\sun$. 
In a basis of $E$ obtained by joining bases of the invariant subspaces $\img{\proj_i}$ (which due to the hermiticity of $\proj_i$ are orthogonal to each other), any $\sun$ color rotation will be block-diagonal, 
\be
U_{(E)} = 
 \mbox{\fontsize{6}{2}\selectfont $
\begin{psmallmatrix} \ \ 
\ \tikeqbis{\begin{scope}[scale=0.6] \draw (-0.2,-0.2) rectangle (0.2,0.2); \end{scope}} \ & \   & \  & \ \\[2mm]
 \  & \ \tikeqbis{\begin{scope}[scale=1.2] \draw (-0.2,-0.2) rectangle (0.2,0.2); \end{scope}} \ & \ & \  \\[2mm]
 \  & \ & \ \tikeqbis{\begin{scope}[scale=1.6] \draw[line width=0.5mm,dotted,color=black!80,opacity =1]   (-0.2,0.2) -- (0.2,-0.2); \end{scope}} \ & \  \\[2mm]
 \ & \  & \ & \ \tikeqbis{\begin{scope}[scale=4] \draw (-0.2,-0.2) rectangle (0.2,0.2); \end{scope}} \ \ 
\end{psmallmatrix} $} \  \ . 
\ee

If each block cannot be further block-diagonalized, \ie, if the chosen set of projectors is of maximal cardinality, each invariant subspace $\img{\proj_i}$ is said to transform under an irreducible representation (irrep) $R_i$ of $\sun$. The tensor product describing the parton system $\left\{ q\bar{q}g\ldots \right\}$ is decomposed into a sum of irreps: 
\be
\bm 3 \otimes \bm{\bar 3} \otimes \bm{8} \otimes \ldots  \ = \ \mathop{\oplus}_{i=1}^{m} R_i \, .
\ee
In order to determine all irreps (also called multiplets) of a parton system, we need to find a maximal, complete set of Hermitian and mutually orthogonal projectors (constructed from the basic legos). 

To conclude this lesson, let us give the pictorial representation of the $\sun$ generators $T^a(R)$ ($a=1\ldots N^2-1$) in the representation $R$ associated to the projector $\proj_R$ (\ie, acting in the invariant subspace $\img{\proj_R}$), 
\be
\label{generator-R}
T^a(R) \ = \  
\tikeqbis{
\begin{scope}[scale=1.5]
\draw (-1.7,0.4) -- (1,0.4); \draw[qua] (-1.7,0.4) -- (-1.2,0.4); \draw[qua] (0.1,0.4) -- (0.2,0.4);
\draw (-1.7,0.2) -- (1,0.2); \draw[anti] (-1.7,0.2) -- (-1.2,0.2); \draw[anti] (0.1,0.2) -- (0.2,0.2);
\draw[line width=0.4mm,dotted,color=black!80,opacity =1] (-1.45,0.1) -- (-1.45,-0.3); 
\draw[line width=0.4mm,dotted,color=black!80,opacity =1] (0.15,0.1) -- (0.15,-0.3); 
\draw[glu] (-1.7,-0.4) -- (1,-0.4) ; 
\draw[glu] (0.5,-1) node[right]{$a$} -- (0.5,-0.5); 
\draw[fill=white,opacity=0.5] (0.5,0.5) .. controls +(-0.15,0) and +(-0.15,0) .. (0.5,-0.5); 
\draw[fill=white,opacity=0.5] (0.5,0.5) .. controls +(0.15,0) and +(0.15,0) .. (0.5,-0.5);
\draw[fill=white,opacity=1] (-0.7,0) circle (0.6); \draw (-0.7,0) node{$\proj_R$} ;
\end{scope}
} \ \ . 
\ee
Indeed, $T^a(R)$ defined in this way is a map of $\img{\proj_R} \to \img{\proj_R}$, and $i \delta \alpha^a \, T^a(R)$ acting on a parton state in $\img{\proj_R}$ is the infinitesimal shift of this state under the infinitesimal color rotation of angles $\delta \alpha^a$. 

\bex
Check pictorially that the $T^a(R)$'s satisfy the $\sun$ Lie algebra. 
\eex

\newpage

\section{Diquark states, Schur's lemma and Casimir charges}
\label{lesson3}

In this lesson, we present a systematic method~\cite{Cvitanovic:2008zz} to find the set of Hermitian projectors on the irreps of a parton system, which can in principle be applied to any parton system. Here it is explained in the very simple case of a $qq$ pair, and other examples will be addressed in lesson~\ref{lesson4}.

\subsection{Irreps of diquark states}
\label{sec:qqirreps} 

Pictorially, a diquark state is represented as
\be
\tikeq{
\draw[qua] (0,0.2) -- (.5,0.2)node[right] {$i$}; 
\draw[fill=white,opacity=1] (-0.1,0.2) circle (0.1); 
\draw[qua] (0,-0.2) -- (.5,-0.2)node[right] {$j$}; 
\draw[fill=white,opacity=1] (-0.1,-0.2) circle (0.1); 
}
\ \equiv \ q^i q^j   \ . 
\ee
As we saw in lesson~\ref{lesson2}, finding a basis of the vector space $\left\{ q^i q^j \right\} \equiv V \otimes V$ (of dimension $N^2$) where all color rotations (represented by $N^2 \times N^2$ matrices) are block-diagonal (and cannot be further block-diagonalized) amounts to finding a maximal and complete set of Hermitian, mutually orthogonal projectors $\proj_i$. 

The case of diquarks being very simple and well known, let us immediately give the result for the relevant set of projectors. It is composed of two projectors corresponding to the symmetrizer and anti-symmetrizer (over the two quark indices), given respectively by:
\be
\label{PSA}
P_{S} = \frac{1}{2} \left( \ 
\tikeq{\draw[qua] (0,0.2) -- (0.8,0.2); \draw[qua] (0,-0.2) -- (0.8,-0.2);}
+
\tikeq{\drawqua(0.3) (0,0.2) -- (0.8,-0.2); \drawqua(0.3) (0,-0.2) -- (0.8,0.2);}
\ \right) 
\ \equiv \ 
\Pplus(1) \ \  ; \ \ \ \ 
P_{A} = \frac{1}{2} \left( \ 
\tikeq{\draw[qua] (0,0.2) -- (0.8,0.2); \draw[qua] (0,-0.2) -- (0.8,-0.2);}
-
\tikeq{\drawqua(0.3) (0,0.2) -- (0.8,-0.2); \drawqua(0.3) (0,-0.2) -- (0.8,0.2);}
\ \right) 
\ \equiv \ \Pminus(1) \ \ . 
\ee
For a system $\left\{ q^i q^j \ldots q^p \right\}$ made up only of quarks, representation theory~\cite{Hamermesh} tells us that the bases of the $\sun$ invariant (and irreducible) subspaces are given by linear combinations of $q^i q^j \ldots q^p$ having different symmetry properties in the permutation of indices. In the present case of two quarks, we can build either a totally symmetric or totally antisymmetric linear combination of $q^i q^j$, leading to the set \eq{PSA} of projectors. 

\bex Verify that $P_{S}$ and $P_{A}$ form a complete set of Hermitian projectors, which are mutually orthogonal. Calculate their ranks. 
\eex

\bex \label{ex:PSPA} $\img{P_{S}}$ and $\img{P_{A}}$ are the subpaces spanned by $V_S^{ij}$ and $V_A^{ij}$ defined by 
\bea
&& V_S^{ij} = (P_{S})^{ij}_{\ \ kl} \, q^k q^l \ = \  
\tikeq{
\draw (0,0.2) -- (1,0.2)node[right] {$i$}; 
\draw[qua] (0.2,0.2) -- (0.3,0.2); 
\draw[qua] (0.8,0.2) -- (0.85,0.2); 
\draw[fill=white,opacity=1] (-0.1,0.2) circle (0.1); 
\draw (0,-0.2) -- (1,-0.2)node[right] {$j$}; 
\draw[qua] (0.2,-0.2) -- (0.3,-0.2); 
\draw[qua] (0.8,-0.2) -- (0.85,-0.2); 
\draw[fill=white,opacity=1] (-0.1,-0.2) circle (0.1); 
\draw[fill=white,opacity=1] (0.4,-0.4) rectangle (0.6,0.4); 
} = \frac{1}{2} (q^i q^j + q^j q^i) \, , \\  
&& V_A^{ij} = (P_{A})^{ij}_{\ \ kl} \, q^k q^l \ = \  
\tikeq{
\draw (0,0.2) -- (1,0.2)node[right] {$i$}; 
\draw[qua] (0.2,0.2) -- (0.3,0.2); 
\draw[qua] (0.8,0.2) -- (0.85,0.2); 
\draw[fill=white,opacity=1] (-0.1,0.2) circle (0.1); 
\draw (0,-0.2) -- (1,-0.2)node[right] {$j$}; 
\draw[qua] (0.2,-0.2) -- (0.3,-0.2); 
\draw[qua] (0.8,-0.2) -- (0.85,-0.2); 
\draw[fill=white,opacity=1] (-0.1,-0.2) circle (0.1); 
\draw[fill=black,opacity=1] (0.4,-0.4) rectangle (0.6,0.4); 
} 
= \frac{1}{2} (q^i q^j - q^j q^i) \, . 
\eea
Check that $\img{P_{S}}$ and $\img{P_{A}}$ are invariant under $\sun$ (which we know from lesson~\ref{lesson2}), by writing how $V_S^{ij}$ and $V_A^{ij}$ transform under finite color rotations (thus showing that they are $\sun$ tensors of rank 2). 
\eex 

In summary, we have the completeness relation 
\be
\tikeq{\draw[qua] (0,0.2) -- (0.8,0.2); \draw[qua] (0,-0.2) -- (0.8,-0.2);}
\ = \  P_{S} + P_{A}  \ = \  \Pplus(1) \, + \, \Pminus(1) \ \  , 
\ee
and the product of two fundamental representations decomposes into a sum of irreps as 
\be
\label{NtimesN}
N \otimes N  \ = \ \frac{N(N+1)}{2} \  \oplus \ \frac{N(N-1)}{2}  \  , 
\ee
where only the dimensions of the irreps are mentioned. Note that in general, knowing the dimension of an irrep is not sufficient to fully determine the irrep, as illustrated by the following exercise. 

\bex
\label{ex:3times3}
For $N=3$, the relation \eq{NtimesN} reads $\bm 3 \otimes \bm 3 = \bm 6 \oplus \bm 3$, but the generators of the irrep of dimension $3$ acting on the subspace spanned by $V_A^{ij}$ are equivalent to $-(T^a)^* = - ({^t}T^a)$, \ie, $V_A^{ij}$ does not transform under ${\rm SU}(3)$ as a quark, but as an antiquark. (Thus, \eq{NtimesN} is  commonly written as $\bm 3 \otimes \bm 3 = \bm 6 \oplus \bm{\bar 3}$.) Prove this by trading the three independent components of $V_A^{ij}$ for the $3$-vector $B_k \equiv \frac{1}{2} \epsilon_{ijk} V_A^{ij}$ (with $\epsilon_{ijk}$ the Levi-Civita tensor of rank 3) and by evaluating the shift $\delta B_k$ under an infinitesimal color rotation. 
\eex

Let us now suppose that we do not know anything about representation theory, and that we therefore do not know from the start the set of Hermitian projectors. We describe below a systematic method to find them. In the $qq$ case, the method is obviously not the most economical, but an advantage of this method is that it can be applied to any parton system composed of quarks, antiquarks and gluons (as we will see in lesson~\ref{lesson4}). 

\vspace{2mm}
The general procedure consists of three steps:
\bi
\begin{shaded}
\item[(i)] Find the maximal number of linearly independent operators (built from the basic legos, thus being $\sun$ invariant tensors when specifying external indices) mapping the vector space to itself. 
\end{shaded}
\ei
\vspace{-2mm}
In the present case of the vector space $V \otimes V$, an operator (or tensor) of this type can be expressed in terms of graphs of the generic form 
\be
\tikeqbis{
\draw (-0.2,0.5) -- (2.9,0.5); 
\draw (-0.2,-0.5) -- (2.9,-0.5);
\draw[glu] (0.5,0.5) -- (1.5,-0.5);
\draw[glu] (0.5,-0.5) -- (1.5,0.5);
\draw[glu] (0.25,0.5) arc(180:0:.3); 
\draw[glu] (0.7,0.8) node{$\scriptstyle\bullet$} .. controls +(0.5,0.5) and +(0,0.6) .. (1.9,0.5) ;
\drawqua(0.5) (1.7,0) arc (180:0:.25); \draw (1.7,0) arc (-180:0:.25);
\draw[glu] (1.3,-0.25)node{$\scriptstyle\bullet$} -- (1.7,-0.05);
\draw[glu] (2.1,-0.2) -- (2.5,-0.5);
\draw[glu] (2.1,0.2) -- (2.5,0.5); 
\draw[qua] (-0.2,0.5) -- (0.2,0.5); \draw[qua] (-0.2,-0.5) -- (0.2,-0.5); 
\draw[qua] (2.5,0.5) -- (2.9,0.5); \draw[qua] (2.5,-0.5) -- (2.9,-0.5); 
} \ \ . 
\ee
Such graphs can be replaced by (linear combinations of) simpler graphs using the following algorithm.

First, we can get rid of any three-gluon vertex appearing in the graph by using the identity (prove it!) 
\be 
\tikeq{ \draw[glu] (0,0) -- (1,0); 
	 \draw[glu] (0.5,-0.5) -- (0.5,0)node{$\scriptstyle\bullet$};}
\ = \ 2 \left( \ 
\tikeq{ 
\draw[anti] (0.2,0) arc(180:0:.3); \draw (0.2,0) arc(-180:0:.3); 
\draw[glu] (0.5,-0.6) -- (0.5,-0.3); \draw[glu] (-0.2,0) -- (0.2,0);  \draw[glu] (0.8,0) -- (1.2,0); 
}
\ + \ 
 \tikeq{
\draw[qua] (0.2,0) arc(180:0:.3); \draw (0.2,0) arc(-180:0:.3); 
\draw[glu] (0.5,-0.6) -- (0.5,-0.3); \draw[glu] (-0.2,0) -- (0.2,0);  \draw[glu] (0.8,0) -- (1.2,0); 
}	\  \right) \ .
\ee
The graphs then reduce to (linear combinations of) graphs where any internal gluon connects at both ends to quark lines. 

Second, every internal gluon can be removed with the help of the Fierz identity \eq{Fierz-map}. So we end up with graphs with four external quark lines (together with airborne quark loops that simply contribute to an irrelevant global factor $N^\ell$) and without gluons. There are only two ways to connect the four external quark lines, and this proves that there are only two independent tensors mapping $V \otimes V$ to itself, namely, 
\be
\unit \ \equiv \   \tikeq{\draw[qua] (0,0.2) -- (0.8,0.2); \draw[qua] (0,-0.2) -- (0.8,-0.2);} \ \ ; \ \ \ \ 
X \ \equiv \ \tikeq{\drawqua(0.3) (0,0.2) -- (0.8,-0.2); \drawqua(0.3) (0,-0.2) -- (0.8,0.2);}  \ \ . 
\ee

\bi
\begin{shaded}
\item[(ii)] Find the "multiplication table" between these operators, and infer the minimal polynomial and eigenvalues of the most interesting one(s).  
\end{shaded}
\ei
\vspace{-2mm}
The multiplication table of the set $\left\{ \unit, X\right\}$ has only one non-trivial entry, 
\be
X^2 \ = \ 
\tikeq{\drawqua(0.3) (0,0.2) -- (0.8,-0.2); \drawqua(0.3) (0,-0.2) -- (0.8,0.2);} \ 
\tikeq{\drawqua(0.3) (0,0.2) -- (0.8,-0.2); \drawqua(0.3) (0,-0.2) -- (0.8,0.2);} 
\ = \  \tikeq{\draw[qua] (0,0.2) -- (0.8,0.2); \draw[qua] (0,-0.2) -- (0.8,-0.2);} 
\ = \ \unit \, .
\ee
The characteristic equation of the operator $X$ is $X^2-\unit=0$. The minimal polynomial of $X$ is thus $x^2-1=(x-1)(x+1)$, which is split with simple roots. From basic linear algebra, it follows that $X$ can be diagonalized (which is not a surprise since $X$ is clearly Hermitian) and has eigenvalues $\left\{ \lambda_1, \lambda_2 \right\} = \left\{ 1,-1 \right\}$. 

In some basis of $\left\{ q^i q^j \right\} \equiv V \otimes V$, the matrix representation of $X$ thus reads
\be
\label{X-mat}
X = 
 \mbox{\fontsize{14}{2}\selectfont $
\begin{psmallmatrix} 
\, \lambda_1 \! \! & \  & \  & \ & \ & \ \\[0mm]
\ & \tikeqbis{\begin{scope}[scale=1] \draw[line width=0.5mm,dotted,color=black!80,opacity =1]   (-0.1,0.1) -- (0.1,-0.1); \end{scope}} \! & \ & \ & \ & \  \\[0mm]
\  & \  & \lambda_1 \! \! & \ & \ & \ \\[0mm]
\  & \  & \ & \lambda_2 \! \! & \  & \ \\[0mm]
 \ & \  & \ & \ & \tikeqbis{\begin{scope}[scale=0.8] \draw[line width=0.5mm,dotted,color=black!80,opacity =1]   (-0.4,0.4) -- (0.4,-0.4); \end{scope}} \! & \ 
 \\[0mm]
 \  & \  & \ & \  & \  & \lambda_2 \, 
\end{psmallmatrix} $} \  \ . 
\ee

\bi
\begin{shaded}
\item[(iii)] Express the projectors on the corresponding eigenspaces in terms of the $\sun$ invariant tensors. 
\end{shaded}
\ei
\vspace{-2mm}
In the above basis, the projectors $P_{\lambda_1}$ and $P_{\lambda_2}$ on the eigenspaces of $X$ are
\be
\label{qq-proj-diagonal}
P_{\lambda_1} = 
 \mbox{\fontsize{14}{2}\selectfont $
\begin{psmallmatrix} 
\, 1 \! \! & \  & \  & \ & \ & \ \\[0mm]
\ & \tikeqbis{\begin{scope}[scale=1] \draw[line width=0.5mm,dotted,color=black!80,opacity =1]   (-0.1,0.1) -- (0.1,-0.1); \end{scope}} \! & \ & \ & \ & \  \\[0mm]
\  & \  & 1 \! \! & \ & \ & \ \\[0mm]
\  & \  & \ & 0 \! \! & \  & \ \\[0mm]
 \ & \  & \ & \ & \tikeqbis{\begin{scope}[scale=0.8] \draw[line width=0.5mm,dotted,color=black!80,opacity =1]   (-0.4,0.4) -- (0.4,-0.4); \end{scope}} \! & \ 
 \\[0mm]
 \  & \  & \ & \  & \  & 0 \, 
\end{psmallmatrix} $} \ \ ; \ \ \ \ 
P_{\lambda_2} = 
 \mbox{\fontsize{14}{2}\selectfont $
\begin{psmallmatrix} 
\, 0 \! \! & \  & \  & \ & \ & \ \\[0mm]
\ & \tikeqbis{\begin{scope}[scale=1] \draw[line width=0.5mm,dotted,color=black!80,opacity =1]   (-0.1,0.1) -- (0.1,-0.1); \end{scope}} \! & \ & \ & \ & \  \\[0mm]
\  & \  & 0 \! \! & \ & \ & \ \\[0mm]
\  & \  & \ & 1 \! \! & \  & \ \\[0mm]
 \ & \  & \ & \ & \tikeqbis{\begin{scope}[scale=0.8] \draw[line width=0.5mm,dotted,color=black!80,opacity =1]   (-0.4,0.4) -- (0.4,-0.4); \end{scope}} \! & \ 
 \\[0mm]
 \  & \  & \ & \  & \  & 1 \, 
\end{psmallmatrix} $} \ \ . 
\ee
Their explicitly $\sun$ invariant form follows from the identities $X = \lambda_1 P_{\lambda_1} + \lambda_2 P_{\lambda_2}$ and $\unit = P_{\lambda_1} + P_{\lambda_2}$, or directly from a mere observation of the matrix $X$ given in \eq{X-mat},  
\be
P_{\lambda_1} = \frac{X - \lambda_2 \unit}{\lambda_1- \lambda_2} = \frac{1}{2} (\unit + X) = P_{S} \ \ ; \ \ \ \ P_{\lambda_2} = \frac{X - \lambda_1 \unit}{\lambda_2- \lambda_1} = \frac{1}{2} (\unit - X) = P_{A} \ . 
\ee

We thus recover the projectors \eq{PSA} without any prior knowledge of representation theory. Note that in the above derivation, the resulting projectors satisfy all requirements {\it by construction}: they are Hermitian and mutually orthogonal, they form a complete set ($P_{S} + P_{A} = \unit$), and they cannot be reduced into a sum of more $\sun$ invariant projectors (since this would imply that there are more than two independent tensors mapping $V \otimes V$ to itself). Therefore, $P_{S}$ and $P_{A}$ must project onto the irreducible representations of diquark states. 

Let us end this section by two important remarks:
\vspace{-2mm}
\bi
\item[\ding{118}] Obviously, the number of projectors (\ie, the number of irreps) cannot exceed the number of independent tensors determined in step (i), $n_{\rm irreps} \leq n_{\rm tensors}$. For diquarks, we have $n_{\rm irreps} = n_{\rm tensors}$, but for more complicated systems we may have $n_{\rm irreps} < n_{\rm tensors}$. This happens when some of the independent tensors are not Hermitian and therefore cannot contribute to the construction of Hermitian projectors. This will be the case for the $qqq$ system considered in lesson~\ref{lesson4}. 
\item[\ding{118}]  When $n_{\rm irreps} < n_{\rm tensors}$, one might naively think that $n_{\rm irreps}$ coincides with the number of independent tensors that are Hermitian, but this is not the case. Indeed, the projectors are linear combinations of some subset of the independent tensors, and the tensors of this subset are thus linear combinations of the projectors.~Since the projectors are not only Hermitian but also commute between them, the same must be true for the independent tensors of the subset.~We infer that in general, it is the largest subset of {\it commuting} Hermitian operators found among the independent tensors that is used to construct the projectors, and thus determines the number of irreps. 
\ei

\subsection{Schur's lemma}

Consider an invariant tensor $A$ mapping the irreps $R_1$ and $R_2$ of respective vector spaces $W_1$ and $W_2$, 
\be
\label{Schur-tensorA}
A \ = \  
\tikeqbis{
\begin{scope}[scale=1]
\draw[glu] (-2.8,0.4) -- (0,0.4) ; 
\draw[glu] (-2.8,-0.4) -- (0,-0.4) ; 
\draw[line width=0.4mm,dotted,color=black!80,opacity =1] (-2.6,0.25) -- (-2.6,-0.25); 
\draw[line width=0.4mm,dotted,color=black!80,opacity =1] (-1,0.25) -- (-1,-0.25); 
\draw (0,0.4) -- (2.8,0.4) ; \draw[qua] (1,0.4) -- (1.05,0.4) ; \draw[qua] (2.6,0.4) -- (2.65,0.4) ; 
\draw[glu] (0,0.25) -- (2.8,0.25) ; 
\draw (0,-0.4) -- (2.8,-0.4) ; \draw[anti] (1,-0.4) -- (1.05,-0.4) ; \draw[anti] (2.6,-0.4) -- (2.65,-0.4) ; 
\draw[line width=0.4mm,dotted,color=black!80,opacity =1] (2.6,0.1) -- (2.6,-0.25); 
\draw[line width=0.4mm,dotted,color=black!80,opacity =1] (1,0.1) -- (1,-0.25); 
\draw[fill=white,opacity=1] (0,0) circle (0.8); 
\draw[fill=white,opacity=1] (-1.8,0) circle (0.6); \draw (-1.8,0) node{$R_1$} ;
\draw[fill=white,opacity=1] (1.8,0) circle (0.6); \draw (1.8,0) node{$R_2$} ;
\end{scope}
} \ \ . 
\ee
Since $A$ is an invariant tensor we can use color conservation: 
\be
\label{Schur-color-rotation}
i \delta \alpha^a \ 
\tikeqbis{
\begin{scope}[scale=1]
\draw[glu] (-2.7,-1) node[right]{$a$} -- (-2.7,-0.5); 
\draw[fill=white,opacity=0.5] (-2.7,0.5) .. controls +(-0.15,0) and +(-0.15,0) .. (-2.7,-0.5); 
\draw[fill=white,opacity=0.5] (-2.7,0.5) .. controls +(0.15,0) and +(0.15,0) .. (-2.7,-0.5);
\draw[glu] (-3.2,0.4) -- (0,0.4) ; 
\draw[glu] (-3.2,-0.4) -- (0,-0.4) ; 
\draw[line width=0.4mm,dotted,color=black!80,opacity =1] (-3,0.25) -- (-3,-0.25); 
\draw[line width=0.4mm,dotted,color=black!80,opacity =1] (-1,0.25) -- (-1,-0.25); 
\draw (0,0.4) -- (2.8,0.4) ; \draw[qua] (1,0.4) -- (1.05,0.4) ; \draw[qua] (2.6,0.4) -- (2.65,0.4) ; 
\draw[glu] (0,0.25) -- (2.8,0.25) ; 
\draw (0,-0.4) -- (2.8,-0.4) ; \draw[anti] (1,-0.4) -- (1.05,-0.4) ; \draw[anti] (2.6,-0.4) -- (2.65,-0.4) ; 
\draw[line width=0.4mm,dotted,color=black!80,opacity =1] (2.6,0.1) -- (2.6,-0.25); 
\draw[line width=0.4mm,dotted,color=black!80,opacity =1] (1,0.1) -- (1,-0.25); 
\draw[fill=white,opacity=1] (0,0) circle (0.8); 
\draw[fill=white,opacity=1] (-1.8,0) circle (0.6); \draw (-1.8,0) node{$R_1$} ;
\draw[fill=white,opacity=1] (1.8,0) circle (0.6); \draw (1.8,0) node{$R_2$} ;
\end{scope}
} 
\ = \ i \delta \alpha^a \ 
\tikeqbis{
\begin{scope}[scale=1]
\draw[glu] (2.7,-1) node[right]{$a$} -- (2.7,-0.5); 
\draw[fill=white,opacity=0.5] (2.7,0.5) .. controls +(-0.15,0) and +(-0.15,0) .. (2.7,-0.5); 
\draw[fill=white,opacity=0.5] (2.7,0.5) .. controls +(0.15,0) and +(0.15,0) .. (2.7,-0.5);
\draw[glu] (-2.8,0.4) -- (0,0.4) ; 
\draw[glu] (-2.8,-0.4) -- (0,-0.4) ; 
\draw[line width=0.4mm,dotted,color=black!80,opacity =1] (-2.6,0.25) -- (-2.6,-0.25); 
\draw[line width=0.4mm,dotted,color=black!80,opacity =1] (-1,0.25) -- (-1,-0.25); 
\draw (0,0.4) -- (3.2,0.4) ; \draw[qua] (1,0.4) -- (1.05,0.4) ; \draw[qua] (3,0.4) -- (3.1,0.4) ; 
\draw[glu] (0,0.25) -- (3.2,0.25) ; 
\draw (0,-0.4) -- (3.2,-0.4) ; \draw[anti] (1,-0.4) -- (1.05,-0.4) ; \draw[anti] (3,-0.4) -- (3.1,-0.4) ; 
\draw[line width=0.4mm,dotted,color=black!80,opacity =1] (3,0.1) -- (3,-0.25); 
\draw[line width=0.4mm,dotted,color=black!80,opacity =1] (1,0.1) -- (1,-0.25); 
\draw[fill=white,opacity=1] (0,0) circle (0.8); 
\draw[fill=white,opacity=1] (-1.8,0) circle (0.6); \draw (-1.8,0) node{$R_1$} ;
\draw[fill=white,opacity=1] (1.8,0) circle (0.6); \draw (1.8,0) node{$R_2$} ;
\end{scope}
} \ \ . 
\ee
In the l.h.s.~of \eq{Schur-color-rotation}, the infinitesimal color rotation acts in the irrep $R_1$ (as is pictorially obvious, it maps $\img{\proj_{R_1}}$ to itself, see lesson~\ref{lesson2}). The same color rotation acts in the irrep $R_2$ in the r.h.s.~of \eq{Schur-color-rotation}. For finite color rotations, \eq{Schur-color-rotation} thus reads 
\be
\label{Schur-assumption}
\forall \ U(\alpha) \in \sun \, , \quad A \, U_{R_1}(\alpha) \ = \ U_{R_2}(\alpha) \, A  \ . 
\ee

The condition \eq{Schur-assumption} is the starting assumption for stating Schur's lemma, which consists of two parts (see \eg\ Ref.~\cite{Georgi:2000vve} for a proof):
\bi
\item[\ding{118}] if $R_1$ and $R_2$ are inequivalent irreducible representations, then $A=0$. 
\ei
This can be proven by showing that if $A \neq 0$, $A$ must be an invertible square matrix, hence $\exists \, A: \forall \, \alpha, U_{R_1}(\alpha) = A^{-1} U_{R_2}(\alpha) A$, \ie, $R_1$ and $R_2$ are simply related by a change of basis, which is the definition of {\it equivalent} representations. 

Viewing the tensor $A$ represented pictorially in \eq{Schur-tensorA} as the "evolution" of a  parton system, we see that in absence of interaction with external color fields, a parton system may change its composition, but always remains in equivalent irreps. 

\bi
\item[\ding{118}] if $W_1=W_2$ (\ie, the initial and final parton content is the same) and $R_1=R_2\equiv R$, then $A$ is proportional to the identity operator in the irrep $R$ (given by $\unit_{R} = \proj_R$),
\ei
\be
\label{Schur-part2}
\tikeqbis{
\begin{scope}[scale=1]
\draw[glu] (-2.8,0.4) -- (2.8,0.4) ; 
\draw[glu] (-2.8,-0.4) -- (2.8,-0.4) ; 
\draw[line width=0.4mm,dotted,color=black!80,opacity =1] (-2.6,0.25) -- (-2.6,-0.25); 
\draw[line width=0.4mm,dotted,color=black!80,opacity =1] (-1,0.25) -- (-1,-0.25); 
\draw[line width=0.4mm,dotted,color=black!80,opacity =1] (2.6,0.25) -- (2.6,-0.25); 
\draw[line width=0.4mm,dotted,color=black!80,opacity =1] (1,0.25) -- (1,-0.25); 
\draw[fill=white,opacity=1] (0,0) circle (0.8); 
\draw[fill=white,opacity=1] (-1.8,0) circle (0.6); \draw (-1.8,0) node{$R$} ;
\draw[fill=white,opacity=1] (1.8,0) circle (0.6); \draw (1.8,0) node{$R$} ;
\end{scope}
} \ = \ c  \ 
\tikeqbis{
\begin{scope}[scale=1]
\draw[glu] (-1.2,0.4) -- (1.2,0.4) ; 
\draw[glu] (-1.2,-0.4) -- (1.2,-0.4) ; 
\draw[line width=0.4mm,dotted,color=black!80,opacity =1] (-0.9,0.25) -- (-0.9,-0.25); 
\draw[line width=0.4mm,dotted,color=black!80,opacity =1] (0.9,0.25) -- (0.9,-0.25); 
\draw[fill=white,opacity=1] (0,0) circle (0.6);  \draw (0,0) node{$R$} ;
\end{scope}
} 
\ = \ c  \  \proj_R \ . 
\ee
The latter equation can be interpreted as follows. Suppose we prepare an incoming multiplet $R$ and try to mix the basis states of this multiplet with the help of an invertible matrix (represented by the middle blob in the l.h.s.~of \eq{Schur-part2}), which is thus a map of $\img{\proj_R} \to \img{\proj_R}$. Due to Schur's lemma \eq{Schur-part2}, up to an overall factor we get exactly the same states. The basis states of a multiplet are uniquely defined.   

Both parts of Schur's lemma are very important results, useful for simplifying calculations and also for intuition. 

To end this section, let us mention that when $R_1 \neq R_2$, a non-zero tensor $A$ of the form \eq{Schur-tensorA} is called a {\it transition operator} (for the transition $R_1 \to R_2$). 
In this case, Schur's lemma can be reformulated as: 
\bi
\item[\ding{118}] There is a transition operator $A$ between $R_1$ and $R_2 \neq R_1$ {\it if and only if} $R_1$ and $R_2$ are equivalent irreps, and $A$ is then a similarity transformation between the two irreps, $U_{R_1}(\alpha) = A^{-1} U_{R_2}(\alpha) A$. (The reciprocal is trivial: if $R_1$ and $R_2$ are equivalent, there exists a non-zero operator $A$ such that $U_{R_1} = A^{-1} U_{R_2} A$, thus mapping $\img{\proj_{R_1}} \to \img{\proj_{R_2}}$, \ie, there is a transition operator $A$ between $R_1$ and $R_2$.) 
\ei

\bex \label{ex:trans-op} When $R_1$ and $R_2$ are different but equivalent irreps, show that the transition operator between $R_1$ and $R_2$ is uniquely defined (up to an overall factor).  
\eex

We will see examples of transition operators in  lesson~\ref{lesson4}, when discussing the $qqq$ system.

\subsection{Casimir charges}

In lesson~\ref{lesson2} we gave the pictorial expression \eq{generator-R} of $\sun$ generators $T^a(R)$ in the irrep $R$. The Casimir operator in this representation is defined by $T^a(R)T^a(R)$, which from Schur's lemma~\eq{Schur-part2} must be proportional to $\unit_{R} = \proj_R$, with a proportionality coefficient named the Casimir charge $C_R$, 
\be
\label{Casimir-R}
T^a(R)T^a(R) \ = \  
\tikeqbis{
\begin{scope}[scale=1]
\draw (-2.3,0.4) -- (0.9,0.4); \draw[qua] (-1.7,0.4) -- (-1.4,0.4); \draw[qua] (0.1,0.4) -- (0.2,0.4);
\draw (-2.3,0.2) -- (0.9,0.2); \draw[anti] (-1.7,0.2) -- (-1.4,0.2); \draw[anti] (0.1,0.2) -- (0.2,0.2);
\draw[line width=0.4mm,dotted,color=black!80,opacity =1] (-1.55,0.1) -- (-1.55,-0.3); 
\draw[line width=0.4mm,dotted,color=black!80,opacity =1] (0.15,0.1) -- (0.15,-0.3); 
\draw[glu] (-2.3,-0.4) -- (0.9,-0.4) ; 
\draw[glu] (-1.9,0.5) .. controls +(0,0.9) and +(0,0.9) .. (0.5,0.5); 
\draw[fill=white,opacity=0.5] (-1.9,0.5) .. controls +(-0.15,0) and +(-0.15,0) .. (-1.9,-0.5); 
\draw[fill=white,opacity=0.5] (-1.9,0.5) .. controls +(0.15,0) and +(0.15,0) .. (-1.9,-0.5);
\draw[fill=white,opacity=1] (-0.7,0) circle (0.6); \draw (-0.7,0) node{$\proj_R$} ;
\draw[fill=white,opacity=0.5] (0.5,0.5) .. controls +(-0.15,0) and +(-0.15,0) .. (0.5,-0.5); 
\draw[fill=white,opacity=0.5] (0.5,0.5) .. controls +(0.15,0) and +(0.15,0) .. (0.5,-0.5);
\end{scope}
} \ = \  C_R \ 
\tikeqbis{
\begin{scope}[scale=1]
\draw (-1.1,0.4) -- (1.1,0.4); \draw[qua] (-1,0.4) -- (-0.6,0.4); \draw[qua] (0.6,0.4) -- (1,0.4);
\draw (-1.1,0.2) -- (1.1,0.2); \draw[anti] (-1,0.2) -- (-0.6,0.2); \draw[anti] (0.6,0.2) -- (1,0.2);
\draw[line width=0.4mm,dotted,color=black!80,opacity =1] (-0.8,0.1) -- (-0.8,-0.3); 
\draw[line width=0.4mm,dotted,color=black!80,opacity =1] (0.8,0.1) -- (0.8,-0.3); 
\draw[glu] (-1.1,-0.4) -- (1.1,-0.4) ; 
\draw[fill=white,opacity=1] (0,0) circle (0.6); \draw (0,0) node{$\proj_R$} ;
\end{scope}
} \ .
\ee
Unlike the generators $T^a(R)$, the Casimir operator commutes with all $\sun$ transformations. In lesson~\ref{lesson1}, we met the Casimir operators $T^a T^a = C_F \unit_{V}$ and $t^a t^a = C_A \unit_{A}$ in the fundamental (quark) and adjoint (gluon) representations. 

\bex
Show that the global Casimir charge $C_R$ of a color state $R$ of two partons (of individual Casimir charges $C_1$ and $C_2$) is given by
\be
\label{CRpartons12}
C_R = C_1 + C_2 + v_{12}(R) \, , 
\ee
where the color "interaction potential" $v_{12}(R)$ of the parton pair in irrep $R$ is defined by
\be
v_{12}(R) \,  \proj_R \ \equiv \ - 2  \ \
\tikeqbis{ \draw[line width=0.4mm,dotted,color=black!80,opacity =1] (0,0.3) -- (1.2,0.3);
 \draw[line width=0.4mm,dotted,color=black!80,opacity =1] (0,-0.3) -- (1.2,-0.3);
\draw[glu] (0.6,-0.3) -- (0.6,0.3); 
} 
\ \proj_R \ = \ -2 \ 
\tikeq{ 
\begin{scope}[scale=1]
\draw[line width=0.4mm,dotted,color=black!80,opacity =1] (-0.4,0.3) -- (1.8,0.3);
 \draw[line width=0.4mm,dotted,color=black!80,opacity =1] (-0.4,-0.3) -- (1.8,-0.3);
\draw[glu] (0,-0.3) -- (0,0.3); 
\draw[fill=white,opacity=1] (0.8,0) circle (0.5); \draw (0.8,0) node{$\proj_R$} ;
\end{scope}
}  \ . 
\ee
What is the generalization of \eq{CRpartons12} to a system of $n >2$ partons? 
\eex

\bex
Calculate the Casimir charges of the two diquark irreps (associated with the projectors $P_{S}$ and $P_{A}$) as a function of $N$.~In which color state is the color interaction potential attractive? 
\eex

\newpage

\section{Color states of \texorpdfstring{$\bm{q \bar{q}}$}{}, \texorpdfstring{$\bm{qg}$}{} and \texorpdfstring{$\bm{qqq}$}{} systems}
\label{lesson4}

Here we will apply the systematic method described in lesson~\ref{lesson3} for diquark states to a few other systems, namely, $q \bar{q}$ pairs, $qg$ pairs, and finally the $qqq$ system. The latter is the simplest system for which the number of irreps is strictly smaller than the number of independent tensors, thus providing a simple illustration of color transitions between equivalent irreps. 

\subsection{\texorpdfstring{$q \bar{q}$}{}}

As a warming up, let us start with $q\bar{q}$ pairs. The systematic procedure for finding how $\bm 3 \otimes \bm{\bar 3}$ decomposes into a sum of irreps is as follows. 
\bi
\item[(i)] In the same way as was done for $qq$ pairs in lesson~\ref{lesson3}, one can obtain a maximal set of independent tensors mapping $V \otimes \overline{V}$ to itself, for instance: 
\ei
\be
\unit \ \equiv \   \tikeqbis{\draw[qua] (0,0.2) -- (0.8,0.2); \draw[anti] (0,-0.2) -- (0.8,-0.2);} \ \ ; \ \ \ \ 
S \ \equiv \  \tikeqbis{ \draw[qua] (0,-.3) -- ++(0.2,0) -- ++(0,.6) -- ++(-0.2,0);  
 	   \draw[anti] (0.7,-.3) -- ++(-0.2,0) -- ++(0,.6) -- ++(0.2,0); } \ . 
\ee
\bi
\item[(ii)] The multiplication table between these tensors has only one non-trivial entry: 
\ei
\be
S^2 \ = \  \tikeqbis{ \draw[qua] (0,-.3) -- ++(0.2,0) -- ++(0,.6) -- ++(-0.2,0);  
 	   \draw[anti] (0.7,-.3) -- ++(-0.2,0) -- ++(0,.6) -- ++(0.2,0); } \ 
	    \tikeqbis{ \draw[qua] (0,-.3) -- ++(0.2,0) -- ++(0,.6) -- ++(-0.2,0);  
 	   \draw[anti] (0.7,-.3) -- ++(-0.2,0) -- ++(0,.6) -- ++(0.2,0); } 
\ = \ N \  
\tikeq{ \draw[qua] (0,-.3) -- ++(0.2,0) -- ++(0,.6) -- ++(-0.2,0);  
 	   \draw[anti] (0.7,-.3) -- ++(-0.2,0) -- ++(0,.6) -- ++(0.2,0); }
\ = \ N S \, . 
\ee
Thus, $S$ has for minimal polynomial $x^2-Nx$, and for eigenvalues $\left\{ \lambda_1, \lambda_2 \right\} = \left\{N, 0 \right\}$. In some basis of $\left\{ q^i q_j \right\} \equiv V \otimes \overline{V}$, we have $S = {\rm diag}\left( \lambda_1 \ldots \lambda_1,  \lambda_2 \ldots \lambda_2 \right)$. 
\bi
\item[(iii)] The projectors on the eigenspaces of $S$ read
\ei
\vspace{-4mm}
\bea
P_{\lambda_1} = \frac{S - \lambda_2 \unit}{\lambda_1- \lambda_2} = \frac{S}{N} = \frac{1}{N} \ \tikeq{ \draw[qua] (0,-.3) -- ++(0.2,0) -- ++(0,.6) -- ++(-0.2,0);  
 	   \draw[anti] (0.7,-.3) -- ++(-0.2,0) -- ++(0,.6) -- ++(0.2,0); } 
\ \equiv \ \proj_{\bf 1} \, ,  \hskip 2.5cm &&  \\ 
P_{\lambda_2} = \frac{S - \lambda_1 \unit}{\lambda_2- \lambda_1} = \unit - \frac{S}{N} \ = 
 \ \tikeq{\draw[qua] (0,0.2) -- (0.8,0.2); \draw[anti] (0,-0.2) -- (0.8,-0.2);} 
\ - \frac{1}{N} \ 
\tikeq{ \draw[qua] (0,-.3) -- ++(0.2,0) -- ++(0,.6) -- ++(-0.2,0);  
 	   \draw[anti] (0.7,-.3) -- ++(-0.2,0) -- ++(0,.6) -- ++(0.2,0); }
= 2 \ 
\tikeq{ \draw[anti] (0,.3) -- ++(.3,-.3);
   	     \draw[qua] (0,-.3) -- ++(.3,.3);
   	     \draw[glu] (.3,0) -- ++(.7,0);
   	     \draw[anti] (1,0) -- ++(.3,.3);
   	     \draw[qua] (1,0) -- ++(.3,-.3) ; } 
\ \equiv \ \proj_{\bf 8} \ .  \hskip 0mm && 
\eea
We thus recover the projectors of ranks $1$ and $N^2-1$ encountered when proving the Fierz identity \eq{Fierz-map} in lesson~\ref{lesson1} (see Exercise~\ref{ex:fierz-proof}). The small bonus of the systematic derivation is that the invariant subspace $\img{\proj_{\bf 8}}$ is now shown to be irreducible (otherwise one would have found more than two independent tensors in step (i)).

The $q \bar{q}$ system thus decomposes into a sum of irreps as 
\be
\label{qqbar-irreps}
\bm 3 \otimes \bm{\bar 3} = \bm 1 \oplus {\bm 8} \, , 
\ee 
where $\bm 1$ and $\bm 8$ denote the trivial (or singlet) and adjoint representations of $\sun$. 

\bex
Write the linear combinations of $q^i q_j$ spanning the invariant subspaces $\img{\proj_{\bf 1}}$ and $\img{\proj_{\bf 8}}$, and check explicitly that they transform as expected under $\sun$. 
\eex

\subsection{\texorpdfstring{$qg$}{}}

Finding the color states of a quark-gluon pair using the systematic procedure is fairly straightforward. Since you get used to it, several steps of the derivation are left as exercises.  
\bi
\item[(i)] For a maximal set of independent tensors mapping $V \otimes A$ to itself, one can take: 
\ei
\vspace{1mm}
\be
\label{set-IAB}
I \ \equiv \Iqg \ ; \ \ \ \  A \ \equiv \Aqg \ ; \ \ \ \ B \ \equiv \Bqg \ . 
\ee
\bex
Prove it by using the same method as for the $qq$ case (see lesson~\ref{lesson3}), and paying attention to quark loops. 
\eex

\bi
\item[(ii)]  The non-trivial entries of the multiplication table are: 
\ei
\be
\label{qg-mult-table}
A^2 = C_F A \ \ ; \ \ \ \ AB = BA = -\frac{1}{2N} \, A \ \ ; \ \ \ \ B^2 = \frac{1}{4}\,I -\frac{1}{2N}\,A \, . 
\ee
\bex
Derive the relations \eq{qg-mult-table} in a pictorial way. How many irreps can we expect from the set of tensors \eq{set-IAB}, and why? Infer the minimal polynomial of $B$, and explain why there is a basis of $\left\{ q^i g^a \right\} \equiv V \otimes A$ where the operator $B$ is represented by the diagonal matrix 
$B = {\rm diag}\left( \lambda_1 \ldots \lambda_1,  \lambda_2 \ldots \lambda_2, \lambda_3 \ldots \lambda_3 \right)$, with $\left\{ \lambda_1, \lambda_2, \lambda_3 \right\} = \left\{\frac{1}{2}, -\frac{1}{2},-\frac{1}{2N} \right\}$. 
\eex

\bi
\item[(iii)] The projectors on the eigenspaces of $B$ are given by 
\ei
\vspace{-2mm}
\bea
P_{\lambda_1} &=& \frac{1}{2} \ \Iqg - \frac{1}{N+1} \ \Aqg + \Bqg \label{P1qg}  \ , \\[1mm]
P_{\lambda_2} &=& \frac{1}{2} \ \Iqg -\frac{1}{N-1} \ \Aqg - \Bqg \label{P2qg} \ , \\[1mm]
P_{\lambda_3} &=& \frac{1}{C_F} \ \Aqg \ .   \label{P3qg} 
\eea
\bex
Obtain the latter projectors using the formula 
\be
\label{Proj-form}
P_{\lambda_i} = \frac{(B - \lambda_j)(B - \lambda_k)}{(\lambda_i - \lambda_j)(\lambda_i - \lambda_k)} \, ,
\ee
where $P_{\lambda_i}$ is the projector on the eigenspace associated to the eigenvalue $\lambda_i$, and $\lambda_j, \lambda_k$ are the two other eigenvalues. (Eq.~\eq{Proj-form} follows from a mere observation of $B$ in its diagonal matrix form.) 
\eex

By construction, the projectors \eq{P1qg}--\eq{P3qg} form a complete (and maximal) set of Hermitian, mutually orthogonal projectors, which thus project on the irreps of a $qg$ pair. 
\bex
Show that the dimensions $K_{\alpha}$ and Casimir charges $C_{\alpha}$ of the irreps read: 
\bea
K_{\alpha} &=& \{ \mbox{\fontsize{12}{2}\selectfont ${ \frac{\Nc (\Nc +2) (\Nc-1)}{2} , \frac{\Nc (\Nc -2) (\Nc+1)}{2}, \Nc }$} \} \, ,  \\ 
C_{\alpha} &=&  \{ \mbox{\fontsize{12}{2}\selectfont ${\frac{(\Nc +1)(3 \Nc-1)}{2\Nc},  \frac{(\Nc -1)(3 \Nc+1)}{2\Nc}, C_F }$} \} \, .
\eea
\eex

Naming the irreps by their dimensions in the case $N=3$, we thus have
\be
\label{qg-irreps}
\bm{8} \otimes \bm{3} = \bm{15} \oplus \bm{\bar{6}} \oplus \bm{3} \, . 
\ee
The second irrep is denoted as $\bm{\bar{6}}$ because for $N=3$, it is equivalent to the {\it complex conjugate} of the irrep $\bm{6}$ appearing in the diquark case $\bm 3 \otimes \bm 3 = \bm 6 \oplus \bm{\bar 3}$ (see lesson~\ref{lesson3}). 

\bex
Prove the latter statement, by first verifying that the projector $P_{\lambda_2}$ can be rewritten as (for $\sun$)
\be
\vspace{-1mm}
P_{\lambda_2} = 2 \left[ \Iqg - P_{\lambda_3} \right] 
\tikeqbis{
\begin{scope}[yshift=-5]
\draw[qua] (-0.5,0) -- (0.4,0); 
\draw[qua] (0.6,0) -- (1.5,0); 
\draw[qua] (0,0.4) .. controls +(0.15,-0.15) and +(-0.2,0) .. (0.4,0.2); 
\draw[qua] (0.6,0.2) .. controls +(0.2,0) and +(-0.15,-0.15) .. (1,0.4); 
\draw[anti] (0,0.4) .. controls +(0.15,0.15) and +(-0.2,0) .. (0.5,0.6) .. controls +(0.2,0) and +(-0.15,0.15) .. (1,0.4); 
\draw[glu]  (-0.5,0.4) -- (0,0.4); 
\draw[glu]  (1,0.4) -- (1.5,0.4); 
\draw[fill=black,opacity=1] (0.4,-0.2) rectangle (0.6,0.4); 
\end{scope}
} \ , 
\ee
then recalling the result of Exercise~\ref{ex:3times3} for $N=3$, and finally using Schur's lemma. 
\eex

Let us emphasize that for $N >3$, the irreps denoted by their "${\rm SU}(3)$ names" $\bm{6}$ (appearing in $\bm 3 \otimes \bm 3 = \bm 6 \oplus \bm{\bar 3}$) and $\bm{\bar{6}}$ (appearing in $\bm{8} \otimes \bm{3} = \bm{15} \oplus \bm{\bar{6}} \oplus \bm{3}$)  have nothing in common and should be considered separately as defined by their associated projectors.

\subsection{\texorpdfstring{$qqq$}{}}

It is important to study the $qqq$ system once in a lifetime. Indeed, for $N=3$ the decomposition of $qqq$ into multiplets contains the color singlet {\it baryons} of the real world. Moreover, it is a simple case allowing to  discuss transition operators. 

\subsubsection{Decomposition into a sum of irreps}  

We follow the same systematic procedure.  
\vspace{-2mm}
\bi
\item[(i)] Complete set of independent tensors mapping $V \otimes V \otimes V \equiv V^{\otimes 3} \to V^{\otimes 3}$
\ei
\vspace{-2mm}
Using the same algorithm as in previous cases, one finds a set of six tensors with only quark lines, 
\vspace{1mm}
\be
\label{set-qqq}
I \ \equiv \ 
\tikeqbis{
\draw[qua] (0,0.3) -- (0.8,0.3); 
\draw[qua] (0,0) -- (0.8,0); 
\draw[qua] (0,-0.3) -- (0.8,-0.3);
}
\ ; \ \ \ 
X_1 \ \equiv \ 
\tikeqbis{
\draw[qua] (0,0.3) -- (0.8,0.3); 
\drawqua(0.2) (0,0) -- (0.8,-0.3);
\drawqua(0.2) (0,-0.3) -- (0.8,0);
} 
\ ; \ \ \ 
X_2 \ \equiv \ 
\tikeqbis{
\drawqua(0.2) (0,0.3) -- (0.8,-0.3); 
\drawqua(0.2) (0,0) -- (0.8,0);
\drawqua(0.2) (0,-0.3) -- (0.8,0.3);
} 
\ ; \ \ \ 
X_3 \ \equiv \ 
\tikeqbis{
\drawqua(0.2) (0,0.3) -- (0.8,0); 
\drawqua(0.2) (0,0) -- (0.8,0.3);
\draw[qua] (0,-0.3) -- (0.8,-0.3);
} 
\ ; \ \ \ 
\sigma \ \equiv \ 
\tikeqbis{
\drawqua(0.2) (0,0.3) -- (0.8,-0.3); 
\drawqua(0.2) (0,0) -- (0.8,0.3);
\drawqua(0.2) (0,-0.3) -- (0.8,0);
} 
\ ; \ \ \ 
\tau \ \equiv \ 
\tikeqbis{
\drawqua(0.2) (0,0.3) -- (0.8,0); 
\drawqua(0.2) (0,0) -- (0.8,-0.3);
\drawqua(0.2) (0,-0.3) -- (0.8,0.3);
} 
\ \ , 
\ee
which correspond to the elements of the permutation group on three objects $S_3$. 

This is the first example we encounter where some of the independent tensors (here $\sigma$ and $\tau = \sigma^\dagger$) are not Hermitian. The six tensors cannot all be used in the construction of Hermitian projectors, and for the $qqq$ system we must have $n_{\rm irreps} < n_{\rm tensors}$ (see the remarks in the end of section~\ref{sec:qqirreps}). 

\bi
\item[(ii)]  Multiplication table
\ei
\vspace{-2mm}
The multiplication table is that of the group $S_3$. 
\vspace{1mm}
\begin{table}[h]
\renewcommand{\arraystretch}{0.4}
\setlength\tabcolsep{12pt}
\begin{center}
\begin{tabular}{c||c|c|c|c|c|c}
  $\cdot$ & $I$  & $\sigma$  & $\tau$  & $X_1$ & $X_2$ & $X_3$  \\[.2cm]
  \hline 
  \hline &&&&&& \\[0cm]
  $I$ & $I$  & $\sigma$  & $\tau$  & $X_1$ & $X_2$ & $X_3$ \\[.2cm]
  \hline &&&&&& \\
  $\sigma$ & $\sigma$  & $\tau$ & $I$  & $X_2$ & $X_3$ & $X_1$ \\[.2cm]
  \hline &&&&&& \\
  $\tau$ &  $\tau$ & $I$  & $\sigma$  & $X_3$ & $X_1$ & $X_2$ \\[.2cm]
  \hline &&&&&& \\
  $X_1$ & $X_1$ & $X_3$ & $X_2$ & $I$ & $\tau$  & $\sigma$    \\[.2cm]
  \hline &&&&&& \\
  $X_2$ & $X_2$ & $X_1$ & $X_3$ & $\sigma$  & $I$ & $\tau$   \\[.2cm]
  \hline &&&&&& \\
  $X_3$ & $X_3$ & $X_2$ & $X_1$ & $\tau$  & $\sigma$ & $I$  \\[.2cm]
\hline
\end{tabular}
\end{center}
  \caption{\label{tab:mult-qqq} Multiplication table of the set of tensors \eq{set-qqq}. To read this table correctly, note that $\sigma X_2 = X_3$, $X_2 \, \sigma = X_1\ldots$, each entry being easily checked pictorially.} 
\end{table}

In order to determine the $qqq$ irreps and associated projectors, we have to find the maximal set of {\it commuting} Hermitian tensors which can be obtained from the set \eq{set-qqq}. From the above table we easily verify the following points:  
\bi
\item[\ding{100}] The tensors $\sigma + \tau$ and $\Sigma \equiv X_1+X_2+X_3$ are Hermitian and commute with every tensor of the set \eq{set-qqq}. We can thus take the tensors $I$, $\sigma + \tau$, $\Sigma$ in the set we are looking for. 
\item[\ding{100}] The tensors $X_1$, $X_2$, $X_3$ are Hermitian, but do not commute with each other. Thus, only one of the $X_i$'s can be added to the set. Let us choose $X_3$. 
\ei
We therefore choose $\{ I, \sigma + \tau, \Sigma, X_3 \}$ as the maximal set from which we can find four irreps and the associated projectors. Note that the tensors $\{ \sigma - \tau, X_2 - X_1 \}$, which complete the former set to form a basis of six tensors mapping $V^{\otimes 3} \to V^{\otimes 3}$, are not necessary for the construction of irreps. Their interpretation will be given shortly. 

In order to proceed to the explicit construction of projectors, we observe that the operator $\sigma + \tau$ has a simple characteristic equation: 
\be
(\sigma + \tau)^2 = \sigma + \tau + 2 I  \, . 
\ee
Its minimal polynomial is thus $x^2-x-2=(x+1)(x-2)$, and in some basis of $\left\{ q^i q^j q^k \right\} \equiv V^{\otimes 3}$, it is represented by the matrix
\be
\label{sigma-plus-tau}
 \sigma + \tau = 
 \mbox{\fontsize{14}{2}\selectfont $
\begin{psmallmatrix} 
\, -1 \! \! & \  & \  & \ & \ & \ \\[0mm]
\ & \tikeqbis{\begin{scope}[scale=1] \draw[line width=0.5mm,dotted,color=black!80,opacity =1]   (-0.1,0.1) -- (0.1,-0.1); \end{scope}} \! & \ & \ & \ & \  \\[0mm]
\  & \  & \! -1 \! \! & \ & \ & \ \\[0mm]
\  & \  & \ & 2 \! \! & \  & \ \\[0mm]
 \ & \  & \ & \ & \tikeqbis{\begin{scope}[scale=0.8] \draw[line width=0.5mm,dotted,color=black!80,opacity =1]   (-0.4,0.4) -- (0.4,-0.4); \end{scope}} \! & \ 
 \\[0mm]
 \  & \  & \ & \  & \  & 2 \, 
\end{psmallmatrix} $} \  \ . 
\ee

\bi
\item[(iii)]  Projectors on eigenspaces 
\ei
The projectors on the eigenspaces of $\sigma + \tau$ associated to the eigenvalues $-1$ and $2$ read
\bea
P_{(-1)} &=& \frac{1}{3} \left( 2 I - \sigma - \tau \right) \ ,  \\ 
P_{(2)} &=& \frac{1}{3} \left( I + \sigma + \tau \right)  \ . 
\eea
By construction, these projectors are Hermitian and mutually orthogonal invariant tensors, and satisfy the completeness relation $P_{(-1)} + P_{(2)} = I = \unit_{V^{\otimes 3}}$. Thus, they split the vector space $V^{\otimes 3}$ into two $\sun$ invariant subspaces. Since we know there are four {\it irreducible} representations, we continue the splitting of the space by considering the action of another tensor of our set of four, \eg\ $X_3$, in each subspace already found. 

Let us start with the action of $X_3$ in $\img{P_{(-1)}}$. Since $X_3^2 = I \Rightarrow X_3^2 \, P_{(-1)}  = P_{(-1)}$, quite trivially the minimal polynomial of $X_3$ restricted to $\img{P_{(-1)}}$ is $x^2-1$, and in some basis of $\img{P_{(-1)}}$ the operator $X_3$ is of the  form 
\be
\left. X_3 \right|_{\img{P_{(-1)}}} = 
 \mbox{\fontsize{14}{2}\selectfont $
\begin{psmallmatrix} 
\, 1 \! \! & \  & \  & \ & \ & \ \\[0mm]
\ & \tikeqbis{\begin{scope}[scale=1] \draw[line width=0.5mm,dotted,color=black!80,opacity =1]   (-0.1,0.1) -- (0.1,-0.1); \end{scope}} \! & \ & \ & \ & \  \\[0mm]
\  & \  & \! 1 \! \! & \ & \ & \ \\[0mm]
\  & \  & \ & \! \! -1 \! \! & \  & \ \\[0mm]
 \ & \  & \ & \ & \tikeqbis{\begin{scope}[scale=0.8] \draw[line width=0.5mm,dotted,color=black!80,opacity =1]   (-0.4,0.4) -- (0.4,-0.4); \end{scope}} \! & \ 
 \\[0mm]
 \  & \  & \ & \  & \  & -1 \, 
\end{psmallmatrix} $} \ \ . 
\ee
The projectors on the respective eigenspaces are given by
\be
\label{Ppm}
P_{\pm} = \frac{I \pm X_3}{2} \, P_{(-1)} = \frac{1}{6} \left[  2 I - \sigma - \tau \pm \left( 2 X_3 - X_1 -X_2 \right)  \right] \, .  
\ee
Applying the same reasoning in the subspace $\img{P_{(2)}}$, we obtain two other projectors (on the two eigenspaces of $X_3$ restricted to $\img{P_{(2)}}$), 
\be
{\tilde P}_{\pm} = \frac{I \pm X_3}{2} \, P_{(2)} = \frac{1}{6} \left[ I + \sigma + \tau \pm \left(X_1+ X_2 +X_3 \right)  \right] \, .  
\ee
We have thus found four projectors satisfying all requirements, and thus the four irreps. 

The projectors ${\tilde P}_{\pm}$ coincide respectively with the symmetrizer and antisymmetrizer over the three quark indices, defined and denoted pictorially as~\cite{Keppeler:2017kwt}: 
\be
\label{sym-antisym}
{\cal S} \equiv \frac{1}{3!} \sum_{\pi \in S_3} \pi  \ \equiv \ 
\tikeqbis{
\begin{scope}[scale=1.3]
\draw[qua] (0,0.3) -- (0.4,0.3); \drawqua(0.6) (0.6,0.3) -- (1,0.3); 
\draw[qua] (0,0) -- (0.4,0); \drawqua(0.6) (0.6,0) -- (1,0);
\draw[qua] (0,-0.3) -- (0.4,-0.3); \drawqua(0.6) (0.6,-0.3) -- (1,-0.3);
\draw[fill=white,opacity=1] (0.4,-0.5) rectangle (0.6,0.5); 
\end{scope}
}
\ \ ; \ \ \ \ 
{\cal A} = \frac{1}{3!} \sum_{\pi \in S_3} {\rm sign}(\pi) \, \pi   \ \equiv \ 
\tikeqbis{
\begin{scope}[scale=1.3]
\draw[qua] (0,0.3) -- (0.4,0.3); \drawqua(0.6) (0.6,0.3) -- (1,0.3); 
\draw[qua] (0,0) -- (0.4,0); \drawqua(0.6) (0.6,0) -- (1,0);
\draw[qua] (0,-0.3) -- (0.4,-0.3); \drawqua(0.6) (0.6,-0.3) -- (1,-0.3);
\draw[fill=black,opacity=1] (0.4,-0.5) rectangle (0.6,0.5); 
\end{scope}
} \ \ . 
\ee
In \eq{sym-antisym} the sums are over all permutations $\pi$ of $S_3$, and ${\rm sign}(\pi)$ denotes the signature of the permutation (with ${\rm sign}(\pi)=+1$ for $\pi=I, \sigma, \tau$ and ${\rm sign}(\pi)=-1$ for $\pi=X_1,X_2,X_3$). 

\bex
Show that the projectors $P_{\pm}$ given by \eq{Ppm} can also be written as 
\be
P_{+} \ = \ \frac{4}{3} \ 
\tikeqbis{
\begin{scope}[scale=1.3]
\draw (-1,0.3) -- (1,0.3); \draw[qua] (-1,0.3) -- (-0.6,0.3); \drawqua(0.6) (0.6,0.3) -- (1,0.3); 
\draw (-1,0) -- (1,0); \draw[qua] (-1,0) -- (-0.6,0); \drawqua(0.6) (0.6,0) -- (1,0);
\draw (-1,-0.3) -- (1,-0.3); \draw[qua] (-1,-0.3) -- (-0.6,-0.3);  \drawqua(0.6) (0.6,-0.3) -- (1,-0.3);
\draw[qua] (-0.2,0.3) -- (0.2,0.3); \draw[qua] (-0.4,0) -- (-0.1,0); \draw[qua] (0.2,0) -- (0.4,0); 
\draw[fill=white,opacity=1] (-0.6,-0.1) rectangle (-0.4,0.4); 
\draw[fill=white,opacity=1] (0.4,-0.1) rectangle (0.6,0.4); 
\draw[fill=black,opacity=1] (-0.1,-0.4) rectangle (0.1,0.1); 
\end{scope}
}
\ \ ; \ \ \ \ 
P_{-} \ =  \ \frac{4}{3} \ 
\tikeqbis{
\begin{scope}[scale=1.3]
\draw (-1,0.3) -- (1,0.3); \draw[qua] (-1,0.3) -- (-0.6,0.3); \drawqua(0.6) (0.6,0.3) -- (1,0.3); 
\draw (-1,0) -- (1,0); \draw[qua] (-1,0) -- (-0.6,0); \drawqua(0.6) (0.6,0) -- (1,0);
\draw (-1,-0.3) -- (1,-0.3); \draw[qua] (-1,-0.3) -- (-0.6,-0.3);  \drawqua(0.6) (0.6,-0.3) -- (1,-0.3);
\draw[qua] (-0.2,0.3) -- (0.2,0.3); \draw[qua] (-0.4,0) -- (-0.1,0); \draw[qua] (0.2,0) -- (0.4,0); 
\draw[fill=black,opacity=1] (-0.6,-0.1) rectangle (-0.4,0.4); 
\draw[fill=black,opacity=1] (0.4,-0.1) rectangle (0.6,0.4); 
\draw[fill=white,opacity=1] (-0.1,-0.4) rectangle (0.1,0.1); 
\end{scope}
} \ \ ,
\ee
corresponding to mixed symmetries in quark indices.
\eex

\bex
Find the dimensions of the four irreps as a function of $N$. 
\eex

Naming as usual the irreps by their dimensions for $N=3$, the irreps associated with the projectors $\proj_{\alpha} \equiv \{ {\tilde P}_{-}, P_{+}, P_{-}, {\tilde P}_{+} \}$ are thus labelled by $\alpha = \{ \bm{1}, \bm{8}_{\mathsize{7}{+}}, \bm{8}_{\mathsize{7}{-}}, \bm{10} \}$. With this notation the decomposition of $\sun$ $qqq$ states into a sum of irreps reads
\be
\bm{3} \otimes \bm{3} \otimes \bm{3} = \bm{1} \oplus \bm{8}_{\mathsize{7}{+}} \oplus \bm{8}_{\mathsize{7}{-}} \oplus \bm{10} \ .
\ee

\bex
For $N=3$, the irrep $R=\bm{1}$ is the representation of the color singlet baryons, and its Casimir must vanish. Calculate the Casimir of the $\sun$ irrep $R=\bm{1}$ for general $N$. 
\eex

\subsubsection{Transition operators}

Among the six independent tensors contributing to the construction of all possible maps of $V^{\otimes 3} \to V^{\otimes 3}$, the set $\{ I, \sigma + \tau, \Sigma, X_3 \}$ (which can be traded for the $\proj_{\alpha}$'s defined just above) allows to build the four irreps. What is the interpretation of the remaining tensors $\sigma - \tau$ and $X_2-X_1$? 

Let us repeat the counting of independent tensors as follows.~Using the completeness relation $\unit = \sum_{\alpha} \proj_{\alpha}$, any $V^{\otimes 3} \to V^{\otimes 3}$ map can be put in the form 
\be
\label{qqq-tensor-counting}
\tikeqbis{
\begin{scope}[scale=1]
\drawqua(0.2) (-1,0.3) -- (0,0.3); \drawqua(0.85)  (0,0.3) -- (1,0.3); 
\drawqua(0.2)  (-1,0) -- (0,0); \drawqua(0.85) (0,0) -- (1,0); 
\drawqua(0.2)  (-1,-0.3) -- (0,-0.3); \drawqua(0.85) (0,-0.3) -- (1,-0.3); 
\draw[fill=white,opacity=1] (0,0) circle (0.6); 
\end{scope}
}
\ = \ \sum_{\alpha, \beta} \ 
\tikeqbis{
\begin{scope}[scale=1]
\drawqua(0.25) (-2.5,0.3) -- (-1.5,0.3); \drawqua(0.5) (-1.5,0.3) -- (0,0.3);
\drawqua(0.55) (0,0.3) -- (1.5,0.3);\drawqua(0.85) (1.5,0.3) -- (2.5,0.3);
\drawqua(0.25) (-2.5,0) -- (-1.5,0); \drawqua(0.5) (-1.5,0) -- (0,0);
\drawqua(0.55) (0,0) -- (1.5,0);\drawqua(0.85) (1.5,0) -- (2.5,0);
\drawqua(0.25) (-2.5,-0.3) -- (-1.5,-0.3); \drawqua(0.5) (-1.5,-0.3) -- (0,-0.3);
\drawqua(0.55) (0,-0.3) -- (1.5,-0.3);\drawqua(0.85) (1.5,-0.3) -- (2.5,-0.3);
\draw[fill=white,opacity=1] (0,0) circle (0.6); 
\draw[fill=white,opacity=1] (-1.5,0) circle (0.5); \draw (-1.5,0) node{$\proj_\alpha$} ;
\draw[fill=white,opacity=1] (1.5,0) circle (0.5); \draw (1.5,0) node{$\proj_\beta$} ;
\end{scope}
} \ \ . 
\ee

From Schur's lemma, in \eq{qqq-tensor-counting} only the terms for which $\alpha$ and $\beta$ are equivalent irreps can contribute. Since equivalent irreps must at least have the same dimension, all possible maps are encompassed by the structures 
\be
\tikeqbis{
\begin{scope}[scale=.8]
\drawqua(0.3) (-2.5,0.3) -- (-1.5,0.3); \drawqua(0.5) (-1.5,0.3) -- (0,0.3);
\drawqua(0.6) (0,0.3) -- (1.5,0.3);\drawqua(0.85) (1.5,0.3) -- (2.5,0.3);
\drawqua(0.3) (-2.5,0) -- (-1.5,0); \drawqua(0.5) (-1.5,0) -- (0,0);
\drawqua(0.6) (0,0) -- (1.5,0);\drawqua(0.85) (1.5,0) -- (2.5,0);
\drawqua(0.3) (-2.5,-0.3) -- (-1.5,-0.3); \drawqua(0.5) (-1.5,-0.3) -- (0,-0.3);
\drawqua(0.6) (0,-0.3) -- (1.5,-0.3);\drawqua(0.85) (1.5,-0.3) -- (2.5,-0.3);
\draw[fill=white,opacity=1] (0,0) circle (0.6); 
\draw[fill=white,opacity=1] (-1.5,0) circle (0.5); \draw (-1.5,0) node{$\alpha$} ;
\draw[fill=white,opacity=1] (1.5,0) circle (0.5); \draw (1.5,0) node{$\alpha$} ;
\end{scope}
} \propto \proj_\alpha \ , \ \ \ \ 
\tikeqbis{
\begin{scope}[scale=.8]
\drawqua(0.3) (-2.5,0.3) -- (-1.5,0.3); \drawqua(0.5) (-1.5,0.3) -- (0,0.3);
\drawqua(0.6) (0,0.3) -- (1.5,0.3);\drawqua(0.85) (1.5,0.3) -- (2.5,0.3);
\drawqua(0.3) (-2.5,0) -- (-1.5,0); \drawqua(0.5) (-1.5,0) -- (0,0);
\drawqua(0.6) (0,0) -- (1.5,0);\drawqua(0.85) (1.5,0) -- (2.5,0);
\drawqua(0.3) (-2.5,-0.3) -- (-1.5,-0.3); \drawqua(0.5) (-1.5,-0.3) -- (0,-0.3);
\drawqua(0.6) (0,-0.3) -- (1.5,-0.3);\drawqua(0.85) (1.5,-0.3) -- (2.5,-0.3);
\draw[fill=white,opacity=1] (0,0) circle (0.6); 
\draw[fill=white,opacity=1] (-1.5,0) circle (0.5); \draw (-1.45,-0.05) node{$\mathsize{11}{\bm{8}}_{\mathsize{4}{+}}$}; 
\draw[fill=white,opacity=1] (1.5,0) circle (0.5); \draw (1.55,-0.05) node{$\mathsize{11}{\bm{8}}_{\mathsize{4}{-}}$};
\end{scope}
} \ \ , \ \ \ \ 
\tikeqbis{
\begin{scope}[scale=.8]
\drawqua(0.3) (-2.5,0.3) -- (-1.5,0.3); \drawqua(0.5) (-1.5,0.3) -- (0,0.3);
\drawqua(0.6) (0,0.3) -- (1.5,0.3);\drawqua(0.85) (1.5,0.3) -- (2.5,0.3);
\drawqua(0.3) (-2.5,0) -- (-1.5,0); \drawqua(0.5) (-1.5,0) -- (0,0);
\drawqua(0.6) (0,0) -- (1.5,0);\drawqua(0.85) (1.5,0) -- (2.5,0);
\drawqua(0.3) (-2.5,-0.3) -- (-1.5,-0.3); \drawqua(0.5) (-1.5,-0.3) -- (0,-0.3);
\drawqua(0.6) (0,-0.3) -- (1.5,-0.3);\drawqua(0.85) (1.5,-0.3) -- (2.5,-0.3);
\draw[fill=white,opacity=1] (0,0) circle (0.6); 
\draw[fill=white,opacity=1] (-1.5,0) circle (0.5); \draw (-1.45,-0.05) node{$\mathsize{11}{\bm{8}}_{\mathsize{4}{-}}$};
\draw[fill=white,opacity=1] (1.5,0) circle (0.5); \draw (1.55,-0.05) node{$\mathsize{11}{\bm{8}}_{\mathsize{4}{+}}$} ;
\end{scope}
} \ \ . 
\ee
We infer that the two latter structures span the same subspace as $\{ \sigma - \tau, X_2-X_1 \} $. In particular, they must be non-zero for some choice of the middle blob, thus defining {\it transition operators} (and proving in passing that $\bm{8}_{\mathsize{7}{+}}$ and $\bm{8}_{\mathsize{7}{-}}$ are equivalent). The operators $\sigma - \tau$ and $X_2-X_1$ are responsible for the transitions $\bm{8}_{\mathsize{7}{+}} \! \leftrightarrow \bm{8}_{\mathsize{7}{-}}$. 

We have seen in lesson~\ref{lesson3} that a transition operator is uniquely defined, see Exercise~\ref{ex:trans-op}. In order to find the $\bm{8}_{\mathsize{7}{+}} \! \leftrightarrow \bm{8}_{\mathsize{7}{-}}$ transition operators, we just need to insert in the middle blob some tensor giving a non-zero result. We can choose $\sigma - \tau$, and thus take 
\bea
Q_a \ \equiv \ P_{+} (\sigma - \tau) P_{-} \ = \ 
\tikeqbis{
\begin{scope}[scale=1]
\drawqua(0.25) (-2.5,0.3) -- (-1.5,0.3); \drawqua(0.5) (-1.5,0.3) -- (0,0.3);
\drawqua(0.55) (0,0.3) -- (1.5,0.3);\drawqua(0.85) (1.5,0.3) -- (2.5,0.3);
\drawqua(0.25) (-2.5,0) -- (-1.5,0); \drawqua(0.5) (-1.5,0) -- (0,0);
\drawqua(0.55) (0,0) -- (1.5,0);\drawqua(0.85) (1.5,0) -- (2.5,0);
\drawqua(0.25) (-2.5,-0.3) -- (-1.5,-0.3); \drawqua(0.5) (-1.5,-0.3) -- (0,-0.3);
\drawqua(0.55) (0,-0.3) -- (1.5,-0.3);\drawqua(0.85) (1.5,-0.3) -- (2.5,-0.3);
\draw[fill=white,opacity=1] (0,0) circle (0.6); \draw (0,0) node{$\sigma - \tau$} ;
\draw[fill=white,opacity=1] (-1.5,0) circle (0.5); \draw (-1.5,0) node{$P_{-}$} ;
\draw[fill=white,opacity=1] (1.5,0) circle (0.5); \draw (1.5,0) node{$P_{+}$} ;
\end{scope}
}  \ \ ,  \hskip 1cm &&  \\[2mm]
Q_b \ \equiv \ Q_a^{\dagger} \ = \ P_{-} (\tau - \sigma) P_{+} \ = \ 
\tikeqbis{
\begin{scope}[scale=1]
\drawqua(0.25) (-2.5,0.3) -- (-1.5,0.3); \drawqua(0.5) (-1.5,0.3) -- (0,0.3);
\drawqua(0.55) (0,0.3) -- (1.5,0.3);\drawqua(0.85) (1.5,0.3) -- (2.5,0.3);
\drawqua(0.25) (-2.5,0) -- (-1.5,0); \drawqua(0.5) (-1.5,0) -- (0,0);
\drawqua(0.55) (0,0) -- (1.5,0);\drawqua(0.85) (1.5,0) -- (2.5,0);
\drawqua(0.25) (-2.5,-0.3) -- (-1.5,-0.3); \drawqua(0.5) (-1.5,-0.3) -- (0,-0.3);
\drawqua(0.55) (0,-0.3) -- (1.5,-0.3);\drawqua(0.85) (1.5,-0.3) -- (2.5,-0.3);
\draw[fill=white,opacity=1] (0,0) circle (0.6); \draw (0,0) node{$\tau - \sigma$} ;
\draw[fill=white,opacity=1] (-1.5,0) circle (0.5); \draw (-1.5,0) node{$P_{+}$} ;
\draw[fill=white,opacity=1] (1.5,0) circle (0.5); \draw (1.5,0) node{$P_{-}$} ;
\end{scope}
} \ \ ,   \hskip 1cm &&
\eea
for the $\bm{8}_{\mathsize{7}{-}} \! \to \bm{8}_{\mathsize{7}{+}}$ and $\bm{8}_{\mathsize{7}{+}} \! \to \bm{8}_{\mathsize{7}{-}}$ transition operators, respectively. 
Note that $Q_a$ and $Q_b$ are nilpotent operators ($Q_a^2=P_{+} (\sigma - \tau) P_{-} P_{+} (\sigma - \tau) P_{-} = 0$ since $P_{-} P_{+} =0$), as is the case for any transition operator mapping two irreps of the same vector space. 

\bex
Express $Q_a$ and $Q_b$ as linear combinations of $\sigma - \tau$ and $X_2-X_1$. 
\eex

\bex 
Check that $Q_a Q_b = 3 P_{+}$ and $Q_b Q_a = 3 P_{-}$, \ie, $Q_a Q_b$ and $Q_b Q_a$ are proportional to the identity operators in the subspaces $\img{P_{+}}$ and $\img{P_{-}}$, respectively. 
\eex

\bex
Find all {\it unitary} similarity transformations between the irreps $\bm{8}_{\mathsize{7}{-}}$ and $\bm{8}_{\mathsize{7}{+}}$. 
\eex

\bex
If we had chosen $X_2$ instead of $X_3$ in our set of four commuting Hermitian tensors (being thus $\{ I, \sigma + \tau, \Sigma, X_2 \}$), what would be the expressions of the projectors on the four irreps? Denoting the new decomposition as $\bm{3} \otimes \bm{3} \otimes \bm{3} = \bm{1} \oplus \bm{8}_{\mathsize{7}{+}}^{'} \oplus \bm{8}_{\mathsize{7}{-}}^{'} \oplus \bm{10}$, compare the multiplets in the irreps $\bm{8}_{\mathsize{7}{+}}^{'}$ and $\bm{8}_{\mathsize{7}{-}}^{'}$ to those in the irreps $\bm{8}_{\mathsize{7}{+}}$ and $\bm{8}_{\mathsize{7}{-}}$. (The multiplet of a $uds$ quark system in the $\sun$ irrep $\alpha$ is defined as the linear combinations $(\proj_{\alpha})^{ijk}_{\ \ \ lmn} \, u^l d^m s^n$, similarly to the diquark case, see Exercise~\ref{ex:PSPA}.)
\eex

\vfill 

\section*{Acknowledgments} Many thanks to the organizers of the school for their kind invitation and to all the participants for the friendly and relaxed atmosphere! Thanks to G.~Jackson for his careful reading of these lecture notes. 

\newpage 

\nocite{*}
\providecommand{\href}[2]{#2}\begingroup\raggedright\endgroup

\end{document}

%% file: macro.tex
\def\ed{\end{document}}
\def\bc{\begin{center}}
\def\ec{\end{center}}
\def\bi{\begin{itemize}}
\def\ei{\end{itemize}}
\def\be{\begin{equation}}
\def\ee{\end{equation}}
\def\bea{\begin{eqnarray}}
\def\eea{\end{eqnarray}}
\newcommand{\nn}{\nonumber} 
\newcommand{\ie}{{i.e.}}  
\newcommand{\eg}{{e.g.}}
\def\bex{\begin{exer}}
\def\eex{\end{exer}}

\def\sun{{\rm SU}(N)}

\newcommand{\unit}{{\mathds{1}}} 
\newcommand{\proj}{{\mathds{P}}} 
\newcommand{\Nc}{\ensuremath{N}\xspace}

\newcommand{\eq}[1]{(\ref{#1})}
\newcommand{\com}[2]{\left[{#1},{#2}\right]}

\def\bm#1{\mbox{\boldmath$#1$}}
\newcommand{\tr}{\mathrm{Tr}\,}

\newcommand{\img}[1]{{\rm img}({#1})}

\newcommand{\mathsize}[2]{\mbox{\fontsize{#1}{2}\selectfont $#2$}}

\newcommand{\Idqqbar}{ \draw[qua] (0,-0.3) -- ++(1,0); \draw[anti] (0,.2) -- ++(1,0);}

\def \Iqg  {\ \tikeqbis{
\begin{scope}[yshift=-5]
\draw[qua] (0,0) -- (1,0); 
\draw[glu] (0,0.4) -- (1,0.4);
\end{scope}}\ }

\def \Aqg {\ \tikeqbis{
\begin{scope}[yshift=-5]
\draw[qua] (0,0) -- (1.2,0); 
\draw[glu] (0,0.4) .. controls +(0.3,0) and +(0,0.1) .. (0.35,0); 
\draw[glu] (0.85,0) .. controls +(0,0.1) and +(-0.3,0) .. (1.2,0.4); 
\end{scope}}\ }

\def \Bqg{\ \tikeqbis{
\begin{scope}[yshift=-5]
\draw[qua] (0,0) -- (1.2,0); 
\draw[glu] (0,0.4) .. controls +(0.5,0) and +(0,0.3) ..   (0.85,0); 
\draw[glu] (0.35,0) .. controls +(0,0.3) and +(-0.5,0) .. (1.2,0.4);
\end{scope}}\ }

\def \Pplus (#1) {
\tikeqbis{
\begin{scope}[scale=#1]
\draw (0,0.2) -- (1,0.2); \draw[qua] (0.2,0.2) -- (0.3,0.2); \draw[qua] (0.8,0.2) -- (0.85,0.2); 
\draw (0,-0.2) -- (1,-0.2); \draw[qua] (0.2,-0.2) -- (0.3,-0.2); \draw[qua] (0.8,-0.2) -- (0.85,-0.2); 
\draw[fill=white,opacity=1] (0.4,-0.4) rectangle (0.6,0.4); 
\end{scope}
}}

\def \Pminus (#1)  {
\tikeqbis{
\begin{scope}[scale=#1]
\draw (0,0.2) -- (1,0.2); \draw[qua] (0.2,0.2) -- (0.3,0.2); \draw[qua] (0.8,0.2) -- (0.85,0.2); 
\draw (0,-0.2) -- (1,-0.2); \draw[qua] (0.2,-0.2) -- (0.3,-0.2); \draw[qua] (0.8,-0.2) -- (0.85,-0.2); 
\draw[fill=black,opacity=1] (0.4,-0.4) rectangle (0.6,0.4); 
\end{scope}
}}